\documentclass[12pt,amsfonts]{article}
\usepackage{amsfonts}
\def\ts{\textstyle}
\def\t{\textstyle}        %DEAR EDITOR:  I PREFER A LARGER, TEXT SIZE FOR `STUFF' IN   $e^{STUFF}$  BUT IF  
\def\one{1\hskip-.37em 1}

\def\half{{\textstyle{\frac{1}{2}}}}
\def\threebytwo{\textstyle{\frac{3}{2}}}
\def\quarter{\textstyle{\frac{1}{4}}}

\def\H{{\cal H}}

\def\threebytwo{\textstyle{\frac{3}{2}}}

\def\p{\phi}

\def\H{{\cal H}}

\def\v{\vskip.3em}

\def\l{\lambda}

\def\F{{\cal F}}

\def\ra{\rightarrow}
\def\tint{{\textstyle\int}}

\def\s{\hskip.08em}

\def\d{\partial}

\def\a{\alpha}
\def\b{\begin{eqnarray*}}  %takes no eqn numbers
\def\e{\end{eqnarray*}}    %takes no eqn numbers
\def\bn{\begin{eqnarray}}  %takes eqn numbers
\def\en{\end{eqnarray}}   %takes eqn numbers
\def\<{\langle}
\def\>{\rangle}

\def\no{\nonumber}

\def\k{\kappa}

\def\{{\lbrace}
\def\}{\rbrace}
\title{Scalar Field Quantization Without \\Divergences
In All Spacetime Dimensions}
\author{John R. Klauder\\
Department of Physics and\\Department of Mathematics\\
University of Florida\\
Gainesville, FL 32611-8440}
\date{ }
\begin{document}
\maketitle
\begin{abstract}Covariant, self-interacting scalar quantum field theories admit solutions for low enough spacetime
dimensions, but when additional divergences appear in higher dimensions, the traditional approach leads to
 results, such as triviality, that are less than satisfactory.
Guided by idealized but soluble {\it non}renormalizable models, a nontraditional proposal for the quantization of covariant scalar
field theories is advanced, which achieves a term-by-term, divergence-free, perturbation analysis of interacting
models expanded about a suitable pseudofree theory, which differs from a free theory by an $O(\hbar^2)$
counterterm. These positive features are realized within a functional integral formulation by a local, nonclassical,
counterterm that effectively transforms  parameter changes in the action from generating mutually singular measures, which are the basis for divergences, to equivalent measures, thereby removing all divergences.
The use of an alternative model about which to perturb is already supported by properties of the classical theory, and is allowed by the inherent ambiguity in the quantization process itself.
 This procedure not only provides acceptable solutions for models for which no acceptable, faithful solution
currently exists, e.g., $\varphi^4_n$,
for spacetime dimensions $n\ge4$, but offers a new, divergence-free solution, for less-singular models as well,
e.g., $\varphi^4_n$, for $n=2,3$. Our analysis implies similar properties for multicomponent scalar
models, such as those associated with the Higgs model.
\end{abstract}

%\pacs{03.65.Db,11.10.-z,11.10.Kk}
\maketitle
\pagenumbering{arabic}
\tableofcontents
\newpage
\section{Introduction}
The standard, textbook procedures for the quantization of covariant scalar quantum fields, such as $\p^4_n$, lead to self-consistent results
for spacetime
dimensions $n=2,3$,  but for spacetime dimensions $n\ge4$,
the same procedures lead to triviality (that is,
to a free or generalized free theory) either rigorously for $n\ge5$ \cite{aiz}, or on the basis of renormalization
group \cite{wilson} and Monte Carlo studies for $n=4$ \cite{wein}. A trivial result implies that the (free) classical model obtained as
the classical limit of the proposed quantum theory does not reproduce the (nonfree) classical  model with which one started,
and, in this case, it is fair to
say that such a quantization is {\it nonfaithful} since its classical limit differs from the original classical theory. This result follows from the natural presumption that the counterterms needed for renormalization of the
quantization process are those suggested by traditional perturbation theory. However, the quantization
process is inherently ambiguous, and it allows for a wide variety of nonclassical terms (proportional to
$\hbar$ so that the right classical limit is formally obtained). Such an ambiguity in quantization may well allow for other, nontraditional counterterms, also proportional to $\hbar$, which lead instead to a nontrivial quantization that may indeed be {\it faithful} in the
sense that the classical limit of the alternative quantum theory leads to the very same classical
theory with which one started. Such an alternative quantization procedure may have a better claim to
be a proper quantization than one that leads to a nonfaithful result.

Some issues need to be addressed immediately: For some problems, a trivial result for a quartic interacting, quantum scalar field is appropriate. For
example, if one is studying a second-order phase transition of some crystalline substance on a cubic lattice of {\it fixed lattice
spacing}, then the quantum formulation of the field at each lattice site is unambiguous, and in that case
the long correlation length makes a self-interacting continuum field a suitable approximating model. Alternatively, when faced with triviality, some authors argue that
nonfaithful quantizations can serve as ``effective theories'', valid in their low order perturbation series, for low energy questions, assuming that some future theory (often identified as Superstring Theory) will resolve high energy issues in a suitable
fashion. While this view is of course possible, it seems more likely that a nonfaithful
quantization is simply the result of an inappropriate quantization procedure, as our discussion will illustrate.

Stated succinctly, our goal is nothing less than finding alternative counterterms that lead to a genuine, faithful quantization of self-interacting scalar fields in all spacetime dimensions; this includes theories that are traditionally viewed as not asymptotically free. Auxiliary fields are {\it not} wanted, and as we shall see they are {\it not} needed. No claim is made that, if successful, these alternative quantization procedures are ``right'' or even ``to be
  preferred''; but they would seem to offer a valid, self-consistent, alternative quantization procedure that  possibly may have some advantages in certain cases.

As we will argue, we indeed are led to a novel, nonclassical (proportional to $\hbar^2$)
counterterm that accomplishes our goal. We could offer the proposed counterterm right away without any
motivating justification, but this would most likely lead to an immediate rejection by the reader, who,
being used to conventional renormalization arguments,
would not appreciate where the alternative counterterm came from and why it should work at all. Instead, we have
chosen a more gentle and indirect line of argument
that not only shows the counterterm's origin but also how and why it works so well.  In only a few words, an outline of that argument reads as follows:
Initially, rather than proceeding from low to high spacetime dimensions,
we choose to go from high to low dimensions. High spacetime dimensions are the realm of traditionally
nonrenormalizable quantum field theories. And to solve the high-dimension story first, we appeal to another
(nonrelativistic) nonrenormalizable---{\it but soluble}---model from which we learn the ``trick'' of how to
choose a suitable counterterm to render such a theory not only soluble but one that admits a finite, term-by-term,  perturbation expansion about a suitably chosen ``pseudofree'' model [similar in this case to a free model plus
an $O(\hbar^2)$ counterterm]; we will also learn why the pseudofree theory and not the usual free theory is the correct theory about which to expand the interaction. After learning what is the trick for the soluble models, we then
apply this knowledge to
high dimension covariant models, and finally we argue for extending this trick
to lower spacetime dimensions as well so as to generate a unified procedure for divergence-free quantization of self-interacting scalar fields in all spacetime dimensions. It is of some interest to report that
preliminary---and as yet rather limited---Monte Carlo data support a {\it nontrivial} behavior for a
covariant $\p^4_4$ model which includes the special counterterm that we propose \cite{stank}.
Finally, in the Appendix we consider the quantum/classical connection for the modified scalar field models based on the use of suitable coherent states.

 Although we focus on scalar fields, similar methods may apply to other
quantum field theories as well. As an example, multiple component scalar fields, as in the Higgs model, are
natural candidates; some brief remarks about this issue appear at the end of the article,
and also at the end of the Appendix. Application to Einstein quantum gravity would also appear to be relevant, especially in the
author's program of Affine Quantum Gravity \cite{affine}; indeed, the present paper may be seen as another contribution to that program.

\subsection{Free and Pseudofree Theories}
We conclude this section with examples of simple systems that clarify the meaning of a pseudofree
theory. Although the counterterm for covariant scalar quantum field theories is proportional to
$\hbar^2$, this is not true in all cases. In particular, the counterterm for the simple, one-dimensional examples we consider in the present section does not involve $\hbar$. For pedagogical reasons, we present this story in {\it two parts}.

{\bf Part 1:} An elementary example of a theory that involves pseudofree behavior is given by
the anharmonic oscillator with the classical action
   \bn A=\tint\{\half[{\dot x}(t)^2-x(t)^2]-g_0\s x(t)^{-4}\}\,dt\;,  \label{eq1}\en
where $g_0\ge0$.
The free theory ($g_0\equiv0$) has solutions $B\cos(t+\beta)$, for general $(B,\beta)$, that freely cross $x=0$. However, when $g_0>0$, {\it no} solution can cross $x=0$, and the limit of interacting solutions as $g_0\ra0$ becomes $\pm\s |B\cos(t+\beta)\s|$.
This latter behavior describes the classical pseudofree model, namely, {\it the classical pseudofree model is that model which is continuously connected to the interacting models as $g_0\ra0$}. In many cases,
the pseudofree theory is the usual free theory, but that is {\it not} the case for this example.

Quantum mechanically, the imaginary-time propagator for the free theory is given by
  \bn  K_f(x'',T;x',0)
  ={\ts\sum}_{n=0}^\infty h_n(x'')\s h_n(x')\,e^{-(n+1/2)T}\;, \en
where $h_n(x)$ denotes the $n$th Hermite function. However, for the interacting quantum theories that follow from (\ref{eq1}), as the coupling $g_0\ra0$, the imaginary-time propagator converges to
  \bn  K_{pf}(x'',T;x',0)=\theta(x''x')\,{\ts\sum}_{n=0}^\infty \s h_n(x'')\,
  [h_n(x')-h_n(-x')\s]\,e^{ -(n+1/2)T}\;,\en
where $\theta(y)=1$ if $y>0$ and $\theta(y)=0$ if $y<0$. Stated otherwise, {\it this imaginary-time propagator
characterizes the quantum pseudofree model as that propagator which is continuously connected to the interacting propagators as $g_0\ra0$}.

The difference between the free and pseudofree propagators has arisen because within a functional integral the interaction acts
partially as a hard core projecting out certain histories that would otherwise be allowed by the free theory; thus the counterterm for this example is represented by that hard core \cite{shlad}. Since the interacting models
are continuously connected to the pseudofree theory as $g_0\ra0$, it is clear that any
perturbation analysis of the interaction term must take place about the pseudofree theory and not
about the free theory.

The potential $x(t)^{-4}$ is clearly singular, and a regularization may be considered useful. Specifically,
let us change $x(t)^{-4}$ to $(|x(t)|+\epsilon)^{-4}$, with $\epsilon>0$. In this case, both classically and
quantum mechanically, the paths can cross the origin (the potential is now {\it bounded!}), and for the so
regularized theory, the limit of the interacting theories as $g_0\ra0$ is the free theory; in that
case, the pseudofree theory is the free theory. If one made a perturbation analysis and could sum up all the
terms {\it exactly}, then the limit as $\epsilon\ra0$ should lead to the discontinuous behavior we have described initially. However, an exact summation of a perturbation series is usually impossible, and without an exact result it is unlikely
that the discontinuous behavior would even be recognized, let alone achieved. Besides the path integral viewpoint, there is an operator viewpoint that also illustrates the same discontinuous behavior. In brief, we are faced with the fact
that the sum $\H=\H_0+g_0\s V$ has the property that $\lim_{g_0\ra0^+}\,[\s\H_0+g_0\s V\s]=\H'_0\not=\H_0$,
which is {\it exactly} what the example above asserts for $\H_0=\half(P^2+Q^2)$, $V=Q^{-4}$, and $\H'_0=\half(P^2|_{D.b.c.}+Q^2)$, with {\it D.b.c.=} Dirichlet boundary conditions \cite{simon}.

It is also interesting to examine two other potentials: $V_2=x^4$ and $V_3=e^{x^4}$. For $V_2$, each term in the perturbation series is finite and the series can be summed (by resummation techniques) to yield the
correct answer; and, of course, that answer passes to the free answer as $g_0\ra0$. For $V_3$, every term in a perturbation expansion in powers of $g_0$ {\it diverges}; nevertheless, the correct answer also passes to the free answer as $g_0\ra0$. Both $V_2$ and $V_3$ represent {\it continuous perturbations} in that
the correct solutions pass to free solutions as $g_0\ra0$. It has been proposed that nonrenormalizable quantum field theories behave in the manner of
$V_3$; instead, we believe that the behavior of nonrenormalizable models is better captured by that of $V=x^{-4}$, that is, as {\it discontinuous perturbations} \cite{science}.

{\bf Part 2:} Let us add some remarks regarding potentials of the form $|x(t)|^{-\a}$,
where $\a>0$. When $\a>2$, the overall classical behavior is the same as for $\a=4$. However, for $\a<2$, the story changes; this change is {\it not} for the solutions to the equations of motion, but {\it the change is for the set of ``variational paths'', i.e., the set of paths that are allowed in the action functional and which therefore form the set of possible paths when deriving the equations of motion}. When $\a<2$, we observe that some variational paths that cross
the axis $x=0$ are actually allowed, while this is not the case for $\a>2$. For example, consider variational paths (or, perhaps, a portion of such paths)
given by $x(t)= \pm|t|^{\beta}$, say, for $-1<t<1$, where $\beta>0$ and the chosen sign is the same as the sign of $t$; in other
words, $x(t)$ is taken as an odd function. In this case, $\dot{x}(t)=\beta|t|^{\beta-1}$, which is square
integrable near $t=0$ provided that $\beta>1/2$. In addition, the interaction term $\tint |x|^{-\a}\,dt<\infty$ provided $\a\,\beta<1$, i.e., $\a<\beta^{-1}<2$. Thus although a finite energy excludes {\it solutions}
of the equations of motion that cross the axis
$x=0$, the {\it set of variational paths allowed by the classical action} includes paths that cross the axis
$x=0$ whenever $\a<2$. Thus, for $\a<2$, it seems that some classical paths allowed by the variational principle are indeed allowed to ``penetrate the barrier'' at $x=0$!

We now officially define the classical pseudofree model by the allowed set of variational paths
for the action functional as $g_0\ra0$, along with the functional form of the action itself,
as compared to the free model which is defined by the allowed set of variational paths when $g_0\equiv0$,
as well as by the form of the free action. For
$\a<2$, some allowed paths cross $x=0$ for both the free and pseudofree models, although these sets are not the same; for example, the
free model allows the variational path $x(t)=t\,\exp(-t^{-2})$, $-1<t<1$, but this path is not allowed for the
pseudofree model for any $\a>0$. The use of the allowed set of variational paths permits a distinction  that is
closer to that regarding the quantum theory than using the set of solutions to the equations of motion. This statement
holds because, for the quantum analysis, it is possible to define the quantum theory for all $\a<2$ so that
the quantum pseudofree theory equals the quantum free theory; how this happens makes for an
 interesting story (especially for $1\le\a<2$), which is presented in \cite{book}. We understand this property as follows: for an imaginary-time,
 functional-integral formulation, the quantum propagator heavily depends on the support of the paths
that contribute to the path integral. The paths that make up that support are in turn determined in some
involved way by what is the exact form of the action functional, and that very same functional form of the action
functional also determines the allowed set of variations that ultimately leads to the equations of motion. Of course, the set of paths that are relevant for the classical theory is disjoint from the set of paths that are
relevant for the functional integral. Nevertheless, both sets of paths are determined by one and the same action
functional! Although, as we have observed, it is possible to arrange that the interacting quantum theory
passes to the usual free theory as $g_0\ra0$  for $\a<2$, this is never possible for $\a>2$. However, this kind of argument is not strong enough to handle the case $\a=2$, which, as it turns out, can be chosen to pass
to the free theory along with those with $\a<2$.

We next turn to field theory where the analysis is more involved. For the classical theory, a hard-core interaction captures the distinction between the free and pseudofree theories, but that image does not apply
too well for the quantum story. As is well known, the support properties in functional integrals for fields is normally such that a change in mass or a change in coupling constant leads to mutually singular measures
signifying that the former and latter support properties are strictly disjoint with probability one. This feature basically accounts for the divergences typically encountered in a perturbation analysis. In our approach, the special counterterm in the field case involves a ``quantum reweighting'', a property similar to typical counterterms in that they evidently reweight the contribution of the field histories and are quantum in that they are proportional to $\hbar$; however, the new counterterm differs from typical counterterms in one fundamental respect: {\it the new counterterm actually renders any parameter change of the original action in the form of {\bf equivalent measures}---and not as mutually singular ones---and this change is the key property  in the elimination of all divergences}. Despite the qualitative differences between the quantum behavior for finitely and infinitely many degrees of freedom, we shall still refer to the ``hard-core interaction'' in the field case whenever the free and pseudofree models differ.

As the initial step in selecting the special counterterm for covariant scalar fields, we analyze how divergences are eliminated in certain {\it soluble} nonrenormalizable models.
For simplicity and clarity, we focus most of our analysis on quartic self-interacting field models;
the analysis for higher self-interaction powers is generally very similar (see, e.g., \cite{book}).

\section{Ultralocal Models}
\subsection{Historical and Contextual Remarks}
The classical action for the quartic ultralocal model is given by
  \bn  A=\int\{\s\half[{\dot\phi}(t,{\bf x})^2-m_0^2\s\phi(t,{\bf x})^2]
  -g_0\s\phi(t,{\bf x})^4\s\}\,dt\s d{\bf x}\;,\label{ulll}\en
  where $g_0\ge0$, ${\dot\phi}(t,{\bf x})=\d\phi(t,{\bf x})/\d t$, ${\bf x}\in{\mathbb R}^s$, and
$1\le s<\infty$. The distinguishing feature of these models is that, with no spatial gradients, the light cone
of covariant models collapses to a temporal line, reflecting the statistical independence of ultralocal fields at any two distinct spatial points.
This model, or some variation of it, has been studied by several authors \cite{ulothers} as a starting point to construct a covariant model by introducing the missing spatial gradient terms, most commonly by some sort of perturbation analysis. These
efforts were approximate in nature in part because the ultralocal model itself was never properly solved as a
genuine starting model. Ultimately, a proper solution of the quantum theory of the ultralocal model itself
was developed \cite{ul}, and several attempts were then made to introduce the gradient terms starting from that proper solution but without any satisfactory outcome. At that point any real efforts to study
covariant models from this standpoint were put on hold.

By themselves, ultralocal models would seem to have no real physical application. In a certain sense that situation changed with two applications when ultralocal models were validly introduced
as an intermediate step in quantizing field theories that formally have a vanishing Hamiltonian but for which
the real dynamics is introduced by imposing one or more constraints. Dirac's approach to the quantization of systems with constraints \cite{dirac} involves complete quantization first before reduction of the
state space by the introduction of any constraints,
and if spatial derivatives only arise from constraints, then this approach properly involves ultralocal models at an intermediate stage.
One such example is Einstein's theory of gravity
which only introduces any spatial derivatives via the constraints, and so an ulralocal quantization can
form part of that study \cite{affine}; another example, involves the use of a reparameterization invariant formulation of a relativistic scalar field \cite{freefield} for which the Hamiltonian vanishes and the dynamics is enforced by means of a constraint.

In the author's work  \cite{ul} on quantizing the model described by (\ref{ulll}), the mathematical solution of interacting ultralocal models was found without introducing any
cutoffs or any perturbation analysis; an even more complete account of that story appears
in \cite{book}.
Although not a physical model by itself, the quantization of the ultralocal model was important since it provided a genuine solution of a truly nonrenormalizable quantum field theory. Only recently has the
mathematical and conceptual ``trick'' been discovered which enables
that particular nonrenormalizable model to avoid divergences and also generate  a nontrivial solution. In the present article, and indeed
 early in our story, we focus on finding that trick and argue that we can exploit that general idea to study other models including covariant models. Doing so gives us an entirely different way to try to introduce the missing
 spatial gradients into the ultralocal models in order to generate a covariant model, an approach
 we again emphasize that preliminary Monte Carlo studies seem to support \cite{stank}.

 After these introductory remarks,
 we take up the study of ultralocal models, their quantization, and discovering the trick leading to
 their solution. In Section 3, we apply that trick to the study of covariant models.

\subsection{Quantization of Ultralocal Models}
Viewed conventionally, it is hard to imagine a quartic interacting field theory that would cause more trouble
in its quantization. On one hand,
 it is clear that ultralocal models are perturbatively nonrenormalizable for any $s\ge1$ simply because the
 imaginary-time, momentum-space propagator is given by $(p_0^2+m_0^2)^{-1}$ rather than by
 $(p_0^2+\Sigma_{j=1}^s p_j^2+m^2_0)^{-1}$ leading to the fact that every closed loop in a conventional diagrammatic
 perturbation analysis
diverges proportional to the volume of spatial-momentum space. On the other
hand, if viewed nonperturbatively, and limited to mass and coupling constant renormalizations,
ultralocal models lead to
free (Gaussian) results simply because when viewed in a lattice regularized formulation,
 the non-Gaussian characteristic function (i.e., the Fourier transform) of the lattice ground-state distribution passes in the continuum limit
 to a Gaussian form based directly on the Central Limit Theorem; a Gaussian ground state leads to a
 free model Hamiltonian and thus to an overall free theory for the continuum quantum theory.
 Clearly, other ideas are required.

Although the quantum theory of these models has been completely solved without introducing cutoffs or without
using perturbation theory \cite{ul,book},
it is pedagogically useful to study the model, at a fixed time, as regularized by a hypercubic spacial lattice with periodic
boundary conditions. If $a>0$ denotes the lattice spacing and $L<\infty$ denotes the number of sites on each
edge, then the ground-state
distribution of the free theory ($g_0\equiv0$) is described (with $M$ a
generic normalization factor) by the characteristic function
\bn   C_f(f)\hskip-1.2em&&=M\int e^{\t  i\Sigma'_k f_k\s \phi_k\s a^s-m_0\Sigma'_k \phi_k^2\s a^s}\,\Pi'_kd\phi_k
\no\\
     &&=e^{\t  -(1/4\s m_0)\Sigma'_k f_k^2\s a^s}\no\\
     &&\ra e^{\t -(1/4\s m_0)\s\int f({\bf x})^2\,d{\bf x}}
\;, \en   where in the last line we have taken the continuum limit, which we define by: $a\ra0$, $L\ra\infty$, such that $a\s L$ is fixed and {\it finite}; the latter restriction allows us to assume that $f({\bf x})$ is an arbitrary
smooth function; in a subsequent limit in which $a\s L\ra\infty$ (generally not considered), the function $f$ needs further specification, such as, $f$ falls to zero at spatial infinity sufficiently fast.
In this expression $k=(k_1,k_2,\ldots,k_s)$, $k_j\in{\mathbb Z}$, labels the sites in this spatial lattice.
Observe: even though the action functional contains the term $\half m^2_0 \p^2$, the ground state wave function involves $\half m_0 \p^2$, in complete analogy to what occurs for a simple harmonic oscillator.

{\bf N. B.} The notation ({\it with primes}) $\Sigma'_k$, $\Pi'_k$, and $N'\equiv L^s$, the number of spatial lattice points, all apply to any single spatial
lattice slice; likewise, the notation ({\it with no primes}) $\Sigma_k$, $\Pi_k$, and $N\equiv  L^sL_0$, ($L_0\ge L$), the number of spacetime lattice points,  all apply to the full spacetime lattice when it comes time to
introduce that finite, $n=s+1$, periodic lattice.

It is of interest to calculate mass-like moments in the ground-state distribution as given  (for $p\ge0$) by
  \bn I_p(m_0)\hskip-1.2em&&\equiv M\int [m_0\Sigma'_k\phi_k^2\s a^s]^p\,e^{\t  -m_0\Sigma'_k \phi_k^2\s a^s}\,\Pi'_kd\phi_k\no\\
     &&=O(N'^p)\;, \label{emoments}\en    
where, once again, $N'\equiv L^s$ is the number of lattice sites in the spatial volume; the approximate evaluation stems from the fact that, when the integrand is expanded out, there are $N'^p$ terms each of which is $O(1)$.
We next wish to develop a perturbation series for the moment $I_1(\widetilde{m}_0)$ in terms of the
expressions $I_p(m_0)$ for $p\ge1$. This series is complicated by the fact that the normalization constant $M$
spends on the mass to a significant degree. In particular, $M(m_0)=(m_0a^s/\pi)^{N'/2}\equiv M_0$, and
similarly $M(\widetilde{m}_0)=(\widetilde{m}_0a^s/\pi )^{N'/2}\equiv \widetilde{M}_0$.
Consequently, a perturbation of the mass, with
$\Delta_0\equiv {\widetilde m}_0-m_0$, leads to
  \bn  &&I_1({\widetilde m}_0)=(\widetilde{M}_0/M_0)\,\{\,I_1(m_0)-(\Delta_0/m_0)\s I_2(m_0)\no\\
            &&\hskip12.2em +\half\s(\Delta_0/m_0)^2\s I_3(m_0)-\cdots\,\}\;, \label{e7}\en
which, apart from the prefactor, and assuming both $m_0$ and $\widetilde{m}_0$ are $O(1)$,  exhibits increasingly divergent contributions in the continuum limit in which $a\ra0$ and $L\ra\infty$
such that $N' a^s=(La)^s$ remains large but finite.

The origin of these divergences can be traced to a
{\it single, specific factor} if we pass
to hyperspherical coordinates, where $\phi_k\equiv\k\s \eta_k$, $\Sigma'_k\phi_k^2\equiv\k^2$, and
$\Sigma'_k\eta_k^2\equiv1$, for which (\ref{emoments}) becomes
   \bn I_p(m_0)=2\s M\int m_0^p\k^{2p}\s a^{sp}\,e^{\t  -m_0\s\k^2\s a^s}\,\k^{(N'-1)}\s d\k\,
   \delta(1-\Sigma'_k
\eta_k^2)\,\Pi'_kd\eta_k\;. \label{ekappa}\en
This expression not only reveals the source of the divergences as the term $N'$ in the measure factor $\k^{(N'-1)}$,
but also reconfirms the approximate evaluation of (\ref{emoments}) by a steepest descent analysis of the $\k$
integration.

Thinking outside the box for a moment, {\it we observe that if we could somehow change the power of $\k$ in the measure of (\ref{ekappa}) to $\k^{(R-1)}$,
where $R$ is a finite factor, then all these divergences would be eliminated!} Even though  this simple remark may seem irrelevant at first sight, this proposal turns out to be {\it exactly} the opening we intend to exploit!

The Gaussian form of the characteristic functional of the ground-state distribution describes the free
ultralocal model. The question naturally arises: what are the candidate functional forms for the characteristic functional of the ground-state distribution for interacting models? Surprisingly, there is much that can be said about this question! The symmetry of the ultralocal model, which implies independent temporal development of the field for each distinct spatial point, implies that any characteristic functional of a ground-state distribution must be of the form
  \bn  C(f)\equiv e^{\t-\tint L[f({\bf x})]\,d{\bf x}}\equiv\int e^{\t i\tint f({\bf x})\s\p({\bf x})\,d{\bf x}}\,\Psi_0(\p)^2\,\Pi'_{\bf x}d\Phi({\bf x})\;. \label{pofirst}\en
The function $L[f]$ fulfills several properties: $L[-f]=L[f]$ (since $\Psi_0(-\p)^2=\Psi_0(\p)^2$ for an
even potential), $L[f]$ is real, and $L[f]\ge0$ (since $|C(f)|\le1$).

Let us now find the most general form for $L[f]$.
For that purpose, choose $f({\bf x})\equiv p$, $-\infty<p<\infty$, ${\bf x}\in \Delta$, and $f({\bf x})\equiv0$ for ${\bf x}\notin\Delta$. Here, $\Delta\subset {\mathbb R}^s$,  and we require that $0<{\bar\Delta}\equiv\tint_\Delta d{\bf x}<\infty$. Thus, as a restricted form of a characteristic functional, it follows that
  \bn e^{\t -{\bar\Delta}L[p]}=\int e^{\t ip\s \l}\,d\mu_\Delta(\l)=\int \cos(p\l)\,d\mu_\Delta(\l) \en
is another characteristic function for some probability measure $\mu_\Delta$ for all $\Delta$. Evidently,
  \bn {\bar\Delta}^{-1}[1-e^{\t-{\bar\Delta}\s L[p]}]=\int[1-\cos(p\l)]\,{\bar\Delta}^{-1}\,d\mu_{\Delta}(\l)\;,\en
  which (assuming convergence) implies that the most general expression for $L[p]$
  is given by
  \bn L[p]\hskip-1.3em&&=\lim_{{\bar\Delta}\ra0}{\bar\Delta}^{-1}[1-e^{\t-{\bar\Delta}\s L[p]}]\no\\
       &&=\lim_{{\bar\Delta}\ra0}\int[1-\cos(p\l)]\,{\bar\Delta}^{-1}\,d\mu_{\Delta}(\l)\no\\
       &&\equiv cp^2+\int[1-\cos(p\l)]\,d\sigma(\l)\;, \en
       where $c\ge0$ and $\sigma$ is a nonnegative measure with the properties
       \bn  \tint[\l^2/(1+\l^2)]\,d\sigma(\l)<\infty\;,\hskip3em \tint d\sigma(\l)\le\infty\;.\label{polast}\en
        {\bf [Remark:} The calculation just presented is part of the theory
       of {\it infinite divisibility} \cite{def} which deals with characteristic functionals $C(f)$
       for which $C(f)^r$ is also a characteristic functional for all $r>0$. The general theory of infinite divisibility also covers those cases where $L[-p]\not=L[p]$, but these more general functional forms are not needed in this article to make our central argument. If $c=0$ above, the resultant
       distributions are generically called Poisson distributions. More specifically: (i) if $d\sigma=A\s\delta(\l-a)\s d\l$,
       $A>0$ and $a\not=0$, (which applies to a case for which $L[-p]\not=L[p]$), then it is called a Poisson distribution; if $\tint d\sigma(\l)<\infty$, then it is called a compound Poisson distribution; and (iii) if $\tint d\sigma(\l)=\infty$, it is called a generalized Poisson distribution.{\bf]}
       The free ultralocal model requires that $c=1/4m_0$ and $\sigma(\l)=0$. Hereafter, we set $c=0$ and focus
       on the second term for which we choose $\sigma$ to be an {\it absolutely continuous measure}, i.e.,
       $d\sigma(\l)\equiv c(\l)^2\,d\l$.

As just demonstrated, besides the Gaussian ground-state distributions,
there are only Poisson ground-state distributions that respect the symmetry of the ultralocal model, and they are described by characteristic functions of the form
  \bn  C(f)=e^{\t -\tint d{\bf x}\s\tint[\s1-\cos(f({\bf x})\s\l)\s]\,c(\l)^2\,d\l}\;, \en
where $\tint[\l^2/(1+\l^2)]\,c(\l)^2\,d\l<\infty$, but $\tint c(\l)^2\,d\l=\infty$ (the latter condition  ensures that the smeared field
operator only has a continuous spectrum;  less obviously, it also ensures a unique ground state \cite{book}). As an
important example, let us assume that $c(\l)^2=b\s\exp(-b\s m\s\l^2)/|\l|$,
where $b$ is a positive constant with dimensions (Length)$^{-s}$, and $m$ is a mass parameter. For this example,
it follows that a suitable modification of the free-field, lattice ground-state distribution leads to
\bn &&\hskip-2em M'\int e^{\t  i\Sigma'_k f_k\phi_k\s a^s -m_0\Sigma'_k\phi_k^2\s a^s}
   \Pi'_k[\s|\phi_k|^{(1-2ba^s)}\s]^{-1}\,\Pi'_kd\phi_k \no \\
    &&\hskip0em =\Pi'_k \{1-(b\s a^s)\tint[\s1-\cos(f_k\s\l)\s]
    \, e^{\t  -b m\l^2}\,d\l/|\l|^{(1-2ba^s)}\}\no\\
   &&\hskip0em \ra e^{\t -b\tint d{\bf x}\s\tint[\s1-\cos(f({\bf x})\s\l)\s]\,
e^{\t  -bm\l^2}\,d\l/|\l|}\;;\label{emean}\en
here we have set $m_0=ba^s m$, $\l=\phi\s\s a^s$, and used the fact that, in the present case,  $M'=(ba^s)^{N'}$ to leading order, which holds because
   \bn (ba^s)\tint e^{\t -bm\l^2}\,d\l/|\l|^{(1-2ba^s)}
   \simeq (2ba^s)\tint_0^B d\l/\l^{(1-2ba^s)}
   =B^{2ba^s}\ra1\;,\en
provided that $0<B<\infty$. {\bf [Remark:} Note well the implicit multiplicative mass renormalization in the
relation $m_0=ba^s  m$.{\bf]}

Observe that the lattice ground-state distribution for this example is
\bn \frac{(b a^s)^{N'}\,e^{\t -m_0\Sigma'_k\phi_k^2\s a^s}}{\Pi'_k|\phi_k|^{(1-2ba^s)}}
  =\frac{(b a^s)^{N'}\,e^{\t -m_0\k^2\s a^s}}{\k^{(N'-2ba^s N')}\,\Pi'_k|\eta_k|^{(1-2ba^s)}}
\label{egsdist}\en
{\it which has exactly the right $\k$-factor to change the $\k$-measure from $\k^{(N'-1)}$ to $\k^{(R-1)}$}, where
in the present case $R=2ba^s N'$ [a finite number chosen in order to ensure a meaningful continuum limit
for (\ref{emean}); more on this choice immediately below]. In addition, to achieve the desired change of the $\k$-measure factor, the symmetry of the ultralocal model ensures that the modification of the free ground-state distribution
to generate an alternative ground-state distribution is unique. If we adopt (\ref{egsdist}) as the
appropriate pseudofree
ground-state distribution, then all divergences due to the integration over $\k$ will disappear!

 Although $R=2ba^sN'$ leads to the proper solution, how could we have initially guessed that this choice would be correct?  Why not consider $R\ra R_1\equiv(2ba^sN')^2$ or
$R\ra R_2\equiv\sqrt{2ba^sN'}$,  both of which are also finite, as candidates for $R$? If we chose $R_1$,
for example, the form of the modification of the ground-state distribution would be
  \bn M''\,e^{\t-m_0\Sigma'_k\p_k^2\,a^s}\,\Pi'_k|\p_k|^{-(1-4b^2a^{2s}N')}\;. \en
  In this case, the modification has an acceptable {\it mathematical} power of $\k$, but that power carries
  the wrong {\it physics}. In particular, the appearance of $N'=L^s$ in the {\it local} modified form of the
  ground-state distribution, $|\p_k|^{-(1-4b^2a^{2s}N')}$, for each site, implies that the {\it local} modification depends on the number of spatial lattice sites $N'$, a parameter quite remote from the local physics. To respect that physics, the form of the local
  ground-state distribution modification, namely $|\p_k|^{-(1-R/N')}$, should be {\it independent} of $N'$,
  and thus $R\propto ba^sN'$ up to a multiple. As it turns out \cite{book}, no physical property depends on that multiple. The multiplier ``2'' is traditional in the authors work, although
  on a few occasions, ``2'' has been replaced by ``1''. {\bf [Remark:} It isn't that $R=R_1$, $R=R_2$, or even $R=1$ (as was the choice in \cite{nbook}) would be wrong {\it for a fixed set of parameters}; rather, those choices are wrong when seeking a {\it proper functional dependence} of $R$ on the parameters. In particular, the choice $R=2ba^sN'$ allows the extension of the spacial volume to infinity in (\ref{emean}) provided that $f({\bf x})$  falls to zero at spatial infinity sufficiently fast for the
  integral to converge; other choices for $R$ generally would not allow a proper extension to spatial
  infinity.{\bf]}

 To illustrate the benefits of the new ground-state distribution (\ref{egsdist}), we offer an example of the advertised lack of divergences. First, note, to leading order, that the normalization factor $M'=(ba^s)^{N'}$ is independent of $m_0$.
Consequently, the perturbation series of a mass-like moment in the new ground-state distribution is given, with ${\widetilde m_0}\equiv m_0+\Delta_0$,  by
    \bn {I}_1({\widetilde m}_0)\hskip-1,2em&&\equiv M'\int [{\widetilde m}_0\Sigma'_k\p_k^2\,a^s] e^{\t-{\widetilde m_0}\Sigma'_k\p_k^2\s a^s }\;\Pi'_k|\p_k|^{-(1-2b\s a^s)}\,\Pi'_k d\p_k \no\\
      &&\hskip0em=({\widetilde m}_0/m_0)M'\sum_{l=0}^\infty\,(-\Delta_0/m_0)^l/l! \int [m_0\Sigma'_k\p_k^2\,a^s]^{l+1}\, e^{\t-{m_0}\Sigma'_k\p_k^2\s a^s }\no\\
       &&\hskip10em\times\Pi'_k|\p_k|^{-(1-2b\s a^s)}\,\Pi'_k d\p_k\no\\
      &&\hskip0em=({\widetilde m}_0/m_0)M'\sum_{l=0}^\infty\,(-\Delta_0/m_0)^l/l!\int [m_0\k^2\,a^s]^{l+1}\, e^{\t-{m_0}\k^2\s a^s}\k^{2ba^sN'-1}\no\\
      &&\hskip10em\times2\,d\k\,\Pi'_k|\eta_k|^{-(1-2b\s a^s)}\,\delta(1-\Sigma'_k\eta_k^2)\,\Pi'_k d\eta_k\no\\
      &&\hskip0em=({\widetilde m}/m)\sum_{l=0}^\infty\,(-\Delta/m)^l/l!\,\Gamma(l+1+\half R)/\Gamma(\half R)\;, \en
      where $\Delta_0= {\widetilde m}_0-m_0$, $\Delta= {\widetilde m}-m$, and $2ba^sN'=R<\infty$. Observe, in this case, that this perturbation series
      is  term-by-term finite unlike the case of a similar expansion (\ref{e7}) for the free ground-state distribution.

      Another example of a perturbation series that is term-by-term finite is given by the following continuum example for the characteristic functional of an interacting ground-state distribution given
      (for $\hbar=1$) by
      \bn C(f)\hskip-1.3em&&=\exp\{-b\tint d{\bf x}\tint[1-\cos(f({\bf x})\s\l)\s]\,e^{\t -mb\l^2-gb^3\s \l^4}\,d\l/|\l|\}\no\\
      &&=\exp\{-b\Sigma_{l=0}^\infty(-gb^3)^l/l!\tint d{\bf x}\tint[1-\cos(f({\bf x})\s\l)\s]\,\l^{4l}\s e^{\t -mb\l^2 }\,d\l/|\l|\}\;.\no\\
      && \label{eq15}\en
    Note well: this expression applies to a particular
    interacting model; specifically, it is {\it not} for a quartic interaction in the classical action, but rather for a model with a sixth power and lesser powers as well \cite{book}.

    In order to identify the important modification represented by $\k^{N'-1}\ra\k^{R-1}$, $R<\infty$, we shall hereafter refer to this procedure as {\it ``measure mashing''}.

\subsection{Role of Sharp Time Averages}
So far we have focussed on sharp time issues for the ultralocal models, and in this section we wish to show how sharp time averages can control spacetime averages of interest as well. Here we let Euclidean time also be periodic with a
lattice spacing of $a$ and $L_0\,(\ge L)$ sites; it is labeled by $k_0\in{\mathbb Z}$. Although we
discuss the present topic in the context of ultralocal models, it has a far wider application, including to the
covariant models that we discuss later. In what follows, $k=(k_0,k_1,k_2,\ldots,k_s)$, and, as already
noted earlier, $\Sigma_k$ $(\Pi_k)$
denotes a spacetime lattice sum (product) while $\Sigma'_k$ $(\Pi'_k)$ denotes a spatial lattice sum (product)
at fixed $k_0$, as was the case previously.

 Let us consider the average of powers of the expression
    \bn \Sigma_{k_0}\s F(\p,a)\,a \;,\en
    where $F(\p,a)$ is a function of lattice points all at a fixed value of $k_0$,
 in any lattice spacetime, based on the standard distribution generated by the exponential of the
 Euclidean action $I$, and which
 we denote by $\<\s(\,\cdot\,)\s\>$.  For example, one may choose $F=m_0^2\Sigma'_k\p_k^2\s a^s$ or $F=g_0\Sigma'_k\p^4_k\s a^s$, etc., and the spacetime average in question is given by
  \bn \<\s[\Sigma_{k_0}\s F(\p,a)\s a\s]^p\s \>\equiv
     M\int [\Sigma_{k_0} F(\p,a)\s a\s]^p\;e^{\t -I(\p,a,\hbar)/\hbar}\;\Pi_k\s d\p_k \;. \en
 We expand the integrand such that
   \bn \<\s[\Sigma_{k_0}\,F(\p,a)\,a]^p\s\>=\Sigma_{k_{0_1},k_{0_2},\ldots,k_{0_p}}\,a^p\,\<\s F(\p_1,a)\,
   F(\p_2,a)\,\cdots\,F(\p_p,a)\s\>\;,\en
   where $\p_j$ here refers to the fact that ``$k_{0_j}=j$" in this term. A straightforward inequality leads to
   \bn &&|\,\<\s F(\p_1,a)\,F(\p_2,a)\,\cdots\,F(\p_p,a)\s\>\,|\no\\
   &&\hskip1cm \le |\,\<\s [F(\p_1,a)]^p\s\>\,\<\s [F(\p_2,a)]^p\s\>\,\cdots\,\<\s [F(\p_p,a)]^p\s\>\,|^{1/p}\;, \en
   which casts the problem into one at sharp time. For sufficiently large (but finite) $L_0$, it follows
   (see below) that this sharp-time expression is given by
      \bn \<\s [F(\p,a)]^p\s\>=\int [F(\p,a)]^p\,\Psi_0(\p)^2\,\Pi'_k\s d\p_k\;, \label{short} \en
      where the integral is taken over fields at a fixed value of $k_0$, $\Psi_0(\p)^2$ denotes the (real) ground-state distribution, and $\Pi'_k$ denotes a product over the spatial lattice at a fixed value of
      $k_0$. Thus we have arrived at the
      important conclusion: {\it if the sharp time average is finite, then the full spacetime average is also finite}.

       This last result has important implications for our claim of a divergence-free perturbation analysis,
      since if we can show that terms in a perturbation series are finite for sharp times, then they are finite for spacetime averages as well. Stated otherwise, if we can show that the sharp time measures have been
      transformed to equivalent measures by establishing a finite perturbation analysis, then it follows that the spacetime measures have also been transformed to equivalent measures.
      
      {\bf [Remark:} For the benefit of
      readers who have forgotten the distinction between such measures, we offer the following one-dimensional
      examples. Let $x\in{\mathbb R}$, and define $dm_1(x)=\exp(-x^2)\s dx$ and $dm_2(x)=\exp(-x^4)\s dx$. Then,
      clearly, there are meaningful functions, $y_{1,2}(x)$ and $y_{2,1}(x)$, such that
      $dm_1(x)=y_{1,2}(x)\,dm_2(x)$ and $dm_2(x)=y_{2,1}(x)\,dm_1(x)$; that is the
      case for two {\it equivalent measures}. Next, define $dm_3(x)=r(x)\s dx$, where $r(x)=1$ for $0<x<1$ and
      $r(x)=0$ otherwise; one says that $m_3$ has support on $[0,1]$. Also define $dm_4(x)=r(x-1)\s dx$, so
      that $m_4$ has support on $[1,2]$. Evidently, there are {\it no} meaningful functions, $y_{3,4}(x)$
      nd $y_{4,3}(x)$,such that
      $dm_3(x)=y_{3,4}(x)\,dm_4(x)$ or $dm_4(x)=y_{4,3}(x)\,dm_3(x)$; one says that $m_3$ and $m_4$
      are {\it mutually singular} since they have disjoint support. Observe that in the case
      of equivalent measures, they may be bridged by a suitable power series, while for
      mutually singular measures, they too could be bridged by a power series, {\it provided they were first
      regularized to become equivalent, but that series would exhibit divergences as the
      regularization was removed!}{\bf]}

      For completeness, we offer the argument that leads to (\ref{short}), which is given
           as follows. Quite generally,
           \bn \<\s F(\p,a)^p\s\>=M{\ts\sum}_{l}\int \<\p|\s l\>\s e^{\t-E_{l}\s T}\s\<l|\s\p\>\s F(\p,a)^p\,\,\Pi'_k\s d\p_k\;, \en
           where $T=L_0\s a$ and we have used the resolution of unity $\one=\tint\s|\s\p\>\<\s\p|\,\Pi'_k\s d\p_k$ for states
           for which ${\hat\p}_k\s|\s\p\>=\p_k\s|\s\p\>$, for all $k$, as well as the eigenvectors $|\s l\>$ and eigenvalues
           $E_{l}$ for which $\H\s|\s l\>=E_{l}\s|\s l\>$.
           For asymptotically large $T$, holding $La$ large but fixed, it follows that only the (unique)
           ground state $|0\>$ contributes, and the
           former expression becomes
              \bn \<\s F(\p,a)^p\s\>=\int \s F(\p,a)^p\s |\<\p|\s 0\>|^2\, \Pi'_k\s d\p_k\;, \en
              now with $M=1$, which is just the expression in (\ref{short}).

\subsection{Lessons from Ultralocal Models}
Observe for the classical ultralocal models that when $g_0>0$ it is necessary that
$\tint\phi(t,{\bf x})^4 dt\s d{\bf x}<\infty$
to derive the equations of motion, but when $g_0=0$ this restriction is absent. Thus the set of allowed variational
paths for $g_0>0$ does {\it not} reduce as $g_0\ra0$ to the set of allowed variational paths for the free theory;
instead, the set of allowed variational paths for $g_0>0$ passes by continuity to a set of allowed variational paths of the
free theory that also incorporates the hard-core consequences of the condition
$\tint\phi(t,{\bf x})^4 dt\s d{\bf x}<\infty$. An interacting classical theory that is not
continuously connected to its own free classical theory is likely to be associated with
an interacting quantum theory that is not continuously connected to its own free quantum theory. This situation is
easy to see for the ultralocal models. The
characteristic function of the ground-state distribution has either a Gaussian or a Poisson form as indicated earlier,
and there is no continuous, reversible path between the two varieties. If one seeks nontriviality, then the
interacting theory must be of the Poisson type; and as the coupling constant vanishes, the continuous limit must
also be a Poisson distribution, namely, the pseudofree model as characterized by (\ref{emean}) \cite{book}.

It is pedagogically useful to compare the story just offered for the two distinct forms---Gaussian {\it or} Poisson---for ultralocal model
characteristic functions with the general form of characteristic functions for even, infinitely divisible, {\it stochastic variables} $X(t)$,
$t\in {\mathbb R}$, which, as we have already implicitly derived before [c.f., Eqs.~(\ref{pofirst}) --
 (\ref{polast})], is given by
  \bn C(f)\hskip-1.2em&&=\<\s e^{\t i\tint f(t)\s X(t)\,dt}\s\>\no\\
    &&=\exp\{-c\tint f(t)^2\,dt -\tint dt\tint [1-\cos(f(t)\l)]\,d\sigma(\l)\}\;, \en
    where $c\ge0$ and $\sigma(\l)$ is a suitable nonnegative measure \cite{def}. This  expression
  implies that $X(t)\equiv X_G(t)+X_P(t)$, namely, that the general answer is the sum of independent
  Gaussian {\it and} Poisson parts. However, in stochastic analysis, it is clear that $X(t)$ is a generalized stochastic process and therefore one is not concerned with its {\it local
  powers} such as $X_{Renormalized}^p(t)$, $p\ge2$. Renormalized local powers for $X_G(t)$ and for $X_P(t)$
  involve distinctly different rules and so they must be considered separately to maintain consistency. Since
  renormalized local field products are absolutely essential for interacting ultralocal models, we have treated
  the Poisson case completely separate from the Gaussian case. Moreover, local powers of a Gaussian ultralocal field, similar to those of Gaussian white noise, are especially difficult to define  and
  in fact have yielded a very limited set of powers \cite{accardi}. The reason this
  issue is of interest to us is because had it been possible to treat the Gaussian and Poisson ultralocal model forms together simultaneously, then there would have
  been a continuous, reversible path between the free model and any interacting model mitigating our simple
  argument that the pseudofree and free models are necessarily distinct. {\bf [Remark:} The reader is familiar
  with the fact that local operators are singular and may be
  concerned that changing the $\k$-measure factor---as we have advocated---may lead to nonsingular operators in
  the continuum limit; as it turns out,
  this is not the case. In fact, consider the continuum limits: $\p_k\equiv{\hat\p}_k\ra{\varphi}({\bf x})$,
  $-i\hbar\s a^{-s}\d/\d\p_k\equiv{\hat\pi}_k\ra{\pi}({\bf x})$, and
  $\half\s [{\hat\pi}_k\s{\hat\p}_k+{\hat\p}_k\s{\hat\pi}_k]\ra{\rho}({\bf x})$. It turns out
  that the continuum field operators formally satisfy canonical commutation relations, $[\s\varphi({\bf x}),{\pi}
  ({\bf y})\s]=i\s\hbar\s\delta({\bf x}-{\bf y})\s\one$, but the quantity
  ${\pi}({\bf x})$, when smeared, is only a form and not an operator [a form requires limitations on bras
  as well as kets,
  as for example the form $\delta(x)$ on the Hilbert space $L^2({\mathbb R})$]. Being only a form severely limits
  the  utility of ${\pi}({\bf x})$. On the other hand,
  \bn [\s{\varphi}({\bf x}),
  {\rho}({\bf y})\s]=i\s\hbar\s\delta({\bf x}-{\bf y})\,{\varphi}({\bf x})\;,\en
   namely, these two fields obey {\it affine commutation relations} (see Appendix) and when smeared lead to
   self-adjoint operators. Thus, although the relevant continuum operators are
noncanonical operators, they are nonetheless ``singular'' in an appropriate sense \cite{book}.
A simple way to see that these operators remain singular is to note the following: before measure mashing,
and with $m_0=O(1)$, the $\k$-measure factor ensures that $\k^2=O(N')$; after measure mashing, and by
choosing $m_0=O(ba^s)$, it again  follows that
$\k^2=O(N')$, but this evaluation comes about by an entirely  different mechanism!{\bf]}

To complete the ultralocal story, we observe that the ground-state distribution for interacting models is also of
the Poisson form, where (with $\hbar=1$)
\bn c(\l)^2=b\,\exp[-y(\l)]/|\l|\en
 for suitable functions $y(\l)$, such as the example in (\ref{eq15}). Each such distribution leads to a lattice
 Hamiltonian
 \bn \H=-\half\s a^{-s}\hbar^2{\ts\sum}'_k\frac{\t\d^2}{\t\d\p_k^2}+\half\s a^{-s}\hbar^2{\ts\sum}'_k\frac{\t 1}{\t\Psi_0(\p)}\s
 \frac{\t\d^2\Psi_0(\p)}{\t\d\p_k^2} \en
 based on the lattice ground state [cf., Eq.~(\ref{emean})]
   \bn \Psi_0(\p)= K\, e^{\t-\half \Sigma'_k\s \s y(\p_k\s a^s)}\,\Pi'_k|\p_k|^{-(1-2ba^s)/2}\;, \label{e32}\en
 where $K$ provides normalization. For a quartic
 interaction, although the explicit function $y$ is unknown,  it is nevertheless chosen so that
 the Hamiltonian has the form (apart from a constant) given by
\bn \H=-\half\s a^{-s}\hbar^2\Sigma'_k{\t\d^2}/{\t\d\p_k^2}+\half m_0^2\Sigma'_k\p_k^2\,a^s
             +g_0\Sigma'_k\p_k^4\,a^s
    +\half\s\hbar^2\,F\Sigma'_k\p_k^{-2}\,a^s\,,\label{e33} \en
    where $F\equiv(\half-ba^s)(\threebytwo-ba^s)\s a^{-2s}$.
{\it The last term in (\ref{e33}) is the sought-for, nonclassical counterterm in the lattice Hamiltonian that effectively leads to a divergence-free formulation!} It is important to observe that not only is the counterterm proportional to $\hbar^2$ but it is also {\it inversely} proportional to the field squared, a fact  that follows directly from two derivatives acting on the denominator factor in (\ref{e32}).  Given the functional form of the ground state, the need for the
given counterterm may also be understood as a required renormalization for the kinetic energy since neither term, by itself, has the ground state in its domain, but only in the given combination.

From the Hamiltonian it is a small step to obtain the lattice
action to be used in a full (Euclidean) lattice spacetime functional integral formulation. Specifically, the lattice action for the quartic ultralocal models is given [for $k\equiv (k_0,k_1,k_2,\ldots,k_s)$ and $k^+\equiv (k_0+1,k_1,k_2,\ldots,k_s)$] by
    \bn I=\half\Sigma_k(\p_{k^+}-\p_k)^2\,\,a^{n-2}+\half m_0^2\Sigma_k\p_k^2\,a^n
             +g_0\Sigma_k\p_k^4\,a^n
    +\half\s\hbar^2\,F\Sigma_k\p_k^{-2}\,a^n\,. \en
    Once again, the last term in the lattice action is the nontraditional, nonclassical counterterm that
    results in a divergence-free formulation.

     Note well, that besides $F$ being
    independent of the lattice size $N'$, as discussed above, $F$ is {\it independent} of the
    coupling constant $g_0$, and as $g_0\ra0$, the counterterm remains unchanged with the result being the
    lattice action for the pseudofree theory, which has the usual form of the action for a free theory augmented
    by a counterterm proportional to $\hbar^2$. From the viewpoint that the counterterm serves to renormalize
    the kinetic energy, its lack of dependence on $g_0$ is perfectly natural. Further analysis \cite{book} shows that, in the continuum limit,  the dynamical spectrum of the pseudofree model has a discrete, {\it uniform} spacing as befits its  name---pseudo{\it free}---and which supports the physical significance of a replacement for the traditional free theory that we have ascribed to it. {\bf [Remark:} This uniform spectral spacing for the pseudofree model serves another purpose. One can imagine alternative proposals for ``pseudofree models'' where instead of replacing $\k^{N'-1}$ by $\k^{R-1}$,
    we could have, for example, replaced $\k^{N'-1}$ by $[1+\cos^2(w\k)]\s\k^{R-1}$. Although incompatible with ultralocal symmetry, this choice would also have eliminated divergences, but at the cost of introducing a new dimensional constant ($w$). The choice of the ``minimal modification'', replacing
    $\k^{N'-1}$ by $\k^{R-1}$, has {\it the virtue of not requiring any new dimensional parameters}. While this
     remark
     is largely academic for the ultralocal models, it has relevance for the covariant models to be discussed later.{\bf]}

    This completes our analysis of the soluble ultralocal models. Notably, our study has found the ``trick'' by
    which such models are solved, and that trick involves choosing a pseudofree model that effectively changes
    the measure factor $\k^{(N'-1)}$ to $\k^{(R-1)}$, where $R=2ba^sN'<\infty$, a process we have already
    agreed to call measure mashing. The number of integration variables is the same before and after measure mashing, only the weighting of the integrand has changed; in particular, one should not confuse measure mashing with dimensional regularization. In hyperspherical variables, the hyperradius $\k$ obeys
    $0\le\k<\infty$, while the $N'$ direction field variables $\eta_k$ each satisfy $-1\le\eta_k\le1$, and
    collectively they fulfill the even more
    restrictive constraint $\Sigma'_k\eta_k^2=1$. Among these variables, only $\k$ has a noncompact
    range, and after measure mashing, integrations over $\k$ no longer yield divergences. Although this
    ``diagnosis'' of the ``disease of divergences'' and its ``prescribed cure'' arose within
    specialized---ultralocal---models, there is no reason not to apply this
    treatment to other models including covariant ones. After all, the volume element expressed in
    hyperspherical coordinates may be regarded as a fundamental ingredient at least as basic as the choice of the  lattice action itself, and changing the effective power of $\k$ can be widely applied as a means to eliminate
    divergences. Observe that measure mashing tames the divergences {\it before} they are encountered, while
    traditional methods cancel
    divergences {\it after} they arise, say from a perturbation study carried out about the free model.
    It is also clear that measure mashing changes the overall measure story, from
    mutually singular to equivalent measures, when parameters of the model change,
     as noted previously. This profound change of measures is
    manifest at sharp times, and thanks to the finiteness of spatial moments implying finiteness of
    spacetime moments, it follows that the full spacetime distribution determined by the lattice action
    augmented by the
    nonclassical counterterm is also such that mutually singular spacetime measures are replaced by
    equivalent spacetime measures.

    As we
    have argued, some potentials are highly singular in such a way that a pseudofree theory different from the
    normal free theory is the one continuously connected to the interacting models, both classically and quantum mechanically. In such cases, a quantum
    perturbation analysis about the free model is inherently incorrect and divergences encountered
    and canceled in such an approach are generally spurious and misleading. In that case, accepting
    the pseudofree model as a different expansion point is already a big first step in the right direction;
    and mashing the measure, which is the principal goal of a suitable choice of the pseudofree
    model, is the second and final step needed to develop a divergence-free quantization formulation. From the
    point of view of the lattice action, the needed change is an $O(\hbar^2)$ counterterm of a kind
    not suggested by conventional perturbation theory (which, after all, is not too surprising since conventional perturbation about the free model in these cases is doomed to failure).

\section{Covariant Models}
We restrict our initial attention to models with the classical
              action given by
\bn A=\tint(\half\s\{{\dot\phi}(x)^2-[{{\nabla}\phi}(x)]^2-m^2_0\s\phi(x)^2\s\} -\l_0\s\phi(x)^4\s)
\,d^n\!x\;,\en
               where $x=(t=x_0,x_1,x_2,\ldots,x_s)\in{\mathbb R}^n$, $n=s+1\ge5$, $\l_0\ge0$, ${\dot\phi}(x)=\d\phi(x)/\d t$,
and $[{\nabla\phi}(x)]^2\equiv\Sigma_{j=1}^s
               (\d\phi(x)/\d x_j)^2$; in this section $\l_0$, and later $\l$, refer to
               coupling constants. It is not obvious, but for the spacetime dimensions in question
               (i.e., $n\ge5$), the
interaction term imposes a restriction on the free action as follows from
               the multiplicative inequality \cite{lady,book}
               \bn \{\tint \phi(x)^4\,d^n\!x\}^{1/2}
               \le C\tint\{{\dot\phi}(x)^2+[\nabla\phi(x)]^2+\phi(x)^2\s\}
\,d^n\!x\;, \en
               where for $n\le4$ (the renormalizable cases), $C=4/3$ is satisfactory, while for
               $n\ge5$ (the nonrenormalizable cases), $C=\infty$ meaning that there are fields for which the
left side diverges while the right side is finite [e.g., $\phi_{singular}(x)=|x|^{-q}\,e^{\t -x^2}$,
               where $n/4\le q< n/2-1$]. {\bf [Remark:} Full disclosure: the inequality
               \bn \{\tint \phi(x)^p\,d^n\!x\}^{2/p}
               \le C\tint\{{\dot\phi}(x)^2+[\nabla\phi(x)]^2+\phi(x)^2\s\}
\,d^n\!x\;, \en
               holds for $n\le 2p/(p-2)$ with $C=4/3$ (the renormalizable cases) and for $n>2p/(p-2)$ with $C=\infty$ (the nonrenormalizable cases).{\bf]}
               As a consequence, for $n\ge5$ the set of variational fields of the interacting
classical theory does {\it not} reduce to the set of variational fields of the free classical theory as the coupling
constant $g_0\ra0$. We now examine the quantum theory in the light of this knowledge, and we initially focus
on choosing a suitable pseudofree model for covariant theories.

 \subsection{Choosing the Covariant Pseudofree Model}
              For covariant scalar fields, the lattice version of a free, nearly massless, quantum theory has a
              characteristic functional for the ground-state distribution given by
              \bn  C_f(f)
              =M'\int e^{\t  i\Sigma'_kf_k\phi_k\s a^s -\Sigma'_{k,l}\phi_k\s A_{k-l}\s\phi_l\s\s
a^{2s}}\,\Pi'_kd\phi_k\:,\en
              where $A_{k-l}$ accounts for the derivatives and a small, well-chosen, artificial mass-like
contribution. The quantum Hamiltonian for this ground state
              (restoring $\hbar$) becomes
              \bn \hskip-.6em\H_f=-\half\hbar^2 a^{-s}{\ts\sum'_k}\s\frac{\t\d^2}{\t\d\phi_k^2}
              +\half{\ts\sum'_{k,l}}
\phi_k\s A^2_{k-l}\s\phi_l\s\s a^{3s}-E_{f} \en
              where $E_{f}$ is a constant ground state energy and
              \bn  A^2_{k-l}\hskip-1.2em&&\equiv {\ts\sum'}_p A_{k-p}\s A_{p-l}\no\\
              &&\hskip0em\equiv\Sigma_{j=1}^s[\s 2\s\delta_{k,l}
-\delta_{k+\delta_j,l}-
              \delta_{k-\delta_j,l}\s]\s a^{-(2s+2)}
               + s\s L^{-2s}\s\delta_{k,l}\s a^{-(2s+2)}\;,\en
              where $k\pm\delta_j\equiv (k_1,k_2,\ldots,k_j\pm1,\ldots,k_s)$, and the last factor is
a small, artificial mass-like term (introduced to deal with the zero mode $\p_k\ra\p_k+\s\xi$). The true mass
term will be introduced later along with the quartic interaction when we discuss the final model.

              We next modify
              the free ground-state distribution in order to suggest a suitable characteristic function
for the pseudofree ground-state
              distribution given by the expression
              \bn &&\hskip-1em C_{pf}(f)
              =M''\int e^{\t  i\Sigma'_kf_k\phi_k\s a^s -\Sigma'_{k,l}\phi_k\s A_{k-l}\s\phi_l\s
\s a^{2s}}\,e^{\t  -W(\phi\s\s a^{(s-1)/2}/\hbar^{1/2})}\no\\
&&\hskip6 em\times  \,
              \{\Pi'_k[\Sigma'_l J_{k,l}\s\phi_l^2]\s\}^{-(1-R/N')/2}\,\Pi'_kd\phi_k\;,
\label{ecovpf}\no\\
\en
              where the constants $J_{k,l}\equiv 1/(2s+1)$ for the $(2s+1)$ points that include $l=k$ and all
              the $2s$  points $l$  that are spatially nearest neighbors to $k\s$; $J_{k,l}\equiv 0$ for all other points. Stated
otherwise,
              the term $\Sigma'_l J_{k,l}\s\phi^2_l$ is {\it an average  of field-squared values} at $k$ and
the $2s$ spatially nearest neighbors to $k$. Note well, that this term leads to a factor of $\k^{-(N'-R)}$ that,
in effect,
              replaces the hyperspherical radius variable measure term $\k^{(N'-1)}$ by the factor $\k^{(R-1)}$
(i.e., mashing the measure), and since $R$ is finite, this choice
              eliminates any divergences caused by integrations over the variable $\k$. Guided by the
              ultralocal models,
              we choose the finite factor $R=2b a^sN'$ in an initial effort to find suitable pseudofree models
              for the covariant theories; we note that this choice leads to acceptable physics regarding the absence of $N'\equiv L^s$ in the local modification that appears in the denominator of the ground state function, and, moreover, support for choosing $R=2b a^sN'$ arises in the analysis leading to Eq.~(\ref{e86}) in the Appendix. The factor $A_{k-l}$ is the same as introduced for the
              free theory, while the function $W$ is implicitly defined below.

              A few general remarks should be made about our proposed choice of the pseudofree model
              for the covariant models in relation to that for the ultralocal models.
              For the ultralocal model, the modification of the ground-state distribution, $\Pi'_k[\p_k^2]^{-(1-2ba^s)/2}$,
              led to the overall measure mashing factor $\k^{-(N'-R)}$, $R=2ba^sN'$,  as required to eliminate
              divergences from integration over the $\k$ variable. The functional form of the modification, as dictated by
              ultralocal symmetry, also leads to incipient divergences from an integration over each $\eta_k$ variable due to a (nearly) non-integrable singularity at $\eta_k=0$. This behavior is proper
              and required for the ultralocal model, but such incipient divergences are {\it not} wanted when we turn attention to the covariant models. For one thing, unlike ultralocal models, neighboring, spatially separated fields are no longer independent in the covariant case, and that fact permits us to redesign the
              local modification to reflect that physics and to do so in such a way that the singularity is {\it integrable} when any subset of the $\eta_k$-variables  vanish. The new modification has taken the form $\Pi'_k[\s\Sigma'_l J_{k,l}\s\p_l^2\s]^{-(1-2ba^s)/2}$, a form
              that still leads to the desired measure mashing factor $\k^{-(N'-R)}$, but leaves the expression $\Pi'_k [\s\Sigma'_l J_{k,l}\eta_l^2\s]^{-(1-2ba^s)/2}$. Due to the property that $\Sigma'_k\eta_k^2=1$, as well as the choice of the coefficients $J_{k,l}$ as described above, it follows that integration over any subset of $\eta_k$ variables over a region where they all may vanish constitutes an
              {\it integrable} singularity, thus avoiding the incipient diverges that arise for the ultralocal case; convergence holds even if we set $2b\s a^s=0$. Integrability follows from the fact that even when $\Upsilon$, $1\le \Upsilon\le N'-1$, of the $\eta_k$ variables are simultaneously passing through zero, there is always less than $\Upsilon$ factors in the denominator that are simultaneously vanishing. As we shall observe below, the finite, local-neighborhood averaging afforded by the choice of $J_{k,l}$ leads, nevertheless, to a {\it local} counterterm in the continuum limit.

              {\bf [Remark:}
              Let us imagine a variable coefficient $\a>0$ (where initially $\a=1$) multiplying all of the spatial lattice
              derivatives in a covariant lattice Hamiltonian. Clearly, the formal limit $\a\ra0$ of the covariant
              model is the ultralocal model. A similar limit of the covariant pseudofree ground-state
              distribution involves a natural change of the matrix denoted by $A_{k-l}$ above to reflect the vanishing spatial derivatives (and perhaps grow a larger mass term) as well as a simultaneous change of the factors $J_{k,l}$ from $1/(2s+1)$ for
              $l=k$ and $l$ equal to any spatially nearest neighbor to $k$, to become simply $J_{k,l}=
              \delta_{k,l}$ appropriate to the ultralocal models. Amusingly, one could also imagine $\a$ as a
              coefficient of (say) {\it just one} of the spatial derivatives (for $s\ge2$) so that as $\a\ra0$, the model has no derivative in one spatial direction but remains covariant in the $s-1$ remaining
              directions. For the pseudofree model, in this case, the original factors $J_{k,l}$
              are changed by such a limit so that the new values are $J_{k,l}=1/(2s-1)$ for $l=k$ and for the nearest neighbors in the $s-1$ spatial directions that remain covariant, while $J_{k,l}=0$ regarding the two nearest neighbors to $k$ in the one direction for
              which the spatial gradient has been removed. Such models have been introduced previously
              under the name {\it covariant diastrophic models} \cite{diastr}, and it would seem that
              our present discussion could also be extended to diastrophic pseudofree models
              as well. To explain the name, ``diastrophic'' is a geological term that refers
              to an extension of the Earth's crust, e.g., as in the generation of a new,
              raised plateau in a formerly flat valley;
              in our usage we imagine that a genuine
              covariant model in $n(=s)$ spacetime dimensions is later extended in an extra spatial dimension
              to a model in $n=s+1$ dimensions in such a way that it has no spatial gradient in the newly added
              spatial direction.{\bf]}

\subsection{The Hamiltonian for the Covariant Pseudofree \\Model}
              The covariant pseudofree Hamiltonian follows from the proposed ground-state wave function $\Psi_{pf}(\p)$
              implicitly contained in (\ref{ecovpf}) in the manner
              \bn \H_{pf}=\half \hbar^2 a^{-s}\s{\ts\sum}'_k\bigg[\s-\s\frac{\t\d^2}{\t\d\p_k^2}+\frac{\t1}{\t\Psi_{pf}(\p)}\,
              \frac{\t\d^2\Psi_{pf}(\p)}{\t\d\p_k^2}\s\bigg]\;. \en
To understand the role
              played by $W$, let us first assume that $W=0$. Then, in taking the necessary second-order
derivatives to derive the potential, there will be a contribution when one derivative acts on the
              $A_{k-l}$ factor in the exponent and the other derivative acts on the denominator factor
              involving $J_{k,l}$. The result will be a cross term that exhibits a long-range interaction
              that would cause difficulty for causality in the continuum limit. Instead, at this point,
              we focus on the Hamiltonian itself as primary (rather than the ground state), and adopt the
              Hamiltonian for the pseudofree model as
               \bn &&\hskip-1em\H_{pf}=-\half\hbar^2\s a^{-s}\s{{\ts\sum}}'_k\frac{\t\d^2}{\t\d\phi_k^2}
              +\half\s{\ts\sum'}_k(\phi_{k^*}-\phi_k)^2\s a^{s-2}\no\\
              &&\hskip1em+\half s(L^{-2s}\s a^{-2}){\ts\sum'}_k\phi_k^2\s a^s
               +\half\s \hbar^2{\ts\sum'}_k{\cal F}_k(\phi)\s a^s-E_{pf}\;,\en
                 where $k^*$ represents a spatially nearest neighbor to $k$ in the positive sense, implicitly
summed over all $s$ spatial directions, and the counterterm ${\cal F}_k(\p)$, which follows from both
                 derivatives acting on the $J_{k,l}$ factor, is given by
                 \bn  &&\hskip-1em{\cal F}_k(\p)
\equiv{\quarter}\s(1-2ba^s)^2\s
          a^{-2s}\s\bigg({\ts\sum'_{\s t}}\s\frac{
  J_{t,\s k}\s \p_k}{[\Sigma'_m\s
  J_{t,\s m}\s\p_m^2]}\bigg)^2\no\\
  &&\hskip2em-{\half}\s(1-2ba^s)
  \s a^{-2s}\s{\ts\sum'_{\s t}}\s\frac{J_{t,\s k}}{[\Sigma'_m\s
  J_{t,\s m}\s\p^2_m]} \no\\
  &&\hskip2em+(1-2ba^s)
  \s a^{-2s}\s{\ts\sum'_{\s t}}\s\frac{J_{t,\s k}^2\s\p_k^2}{[\Sigma'_m\s
  J_{t,\s m}\s\p^2_m]^2}\;. \label{eF} \en
  We observe that this form for the counterterm leads to a local potential in the continuum limit
  even though it is a rather unfamiliar one. {\bf (Remark:} If $J_{k,l}$ is taken as $\delta_{k,l}$, the resultant counterterm is that appropriate to the ultralocal models.{\bf)}

  With this involved counterterm, the pseudofree Hamiltonian is effectively defined, and we choose the implicitly
given expression for the pseudofree ground state to be the ground state $\Psi_{pf}(\p)$ for this Hamiltonian. For large
  $\phi$ values the $A_{k-l}$ term well represents the solution, and for small $\phi$ values the denominator
terms involving $J_{k,l}$ also well represents the solution. The role of $E_{pf}$ and the (unknown) function $W$
is to fine tune the solution so that it satisfies the equation $\H_{pf}\s \Psi_{pf}(\p)=0$. The
manner in which both $a$ and $\hbar$ appear in $\H_{pf}$ dictates how they appear in
  $W$ as $W(\phi\s\s a^{(s-1)/2}/\hbar^{1/2})$.\v
\subsection{Final Form of Lattice Hamiltonian and \\Lattice Action}
  It is but a small step to propose expressions for the lattice Hamiltonian and lattice action in
  the presence of the proper mass term and the quartic interaction. The lattice Hamiltonian is given by
  \bn &&\hskip-1em\H=-\half\hbar^2\s a^{-s}\s{{\ts\sum}}'_k\frac{\t\d^2}{\t\d\phi_k^2}
               +\half\s{\ts\sum'}_k(\phi_{k^*}-\phi_k)^2\s a^{s-2}\no\\
               &&\hskip1.4em +\half s(L^{-2s}\s a^{-2}){\ts\sum'}_k\phi_k^2\s a^s
                +\half m^2_0{\ts\sum}'_k\phi^2_k\s a^s\no\\&&\hskip1.4em +\l_0\s{\ts\sum}'_k\phi^4_k\s a^s
               +\half\s \hbar^2{\ts\sum'}_k{\cal F}_k(\phi)\s a^s-E \;, \en
  and the Euclidean lattice action reads
                   \bn &&\hskip-.8em I(\p,a,\hbar)
               =\half\s{\ts\sum}_k\s{\ts\sum}_{k^*}(\phi_{k^*}-\phi_k)^2\s a^{n-2}
               +\half s(L^{-2s}\s a^{-2})
{\ts\sum}_k\phi_k^2\s a^n\no\\
               &&\hskip6em +\half m^2_0{\ts\sum}_k\phi^2_k\s a^n+\l_0\s{\ts\sum}_k\phi^4_k\s a^n
               +\half\s \hbar^2{\ts\sum}_k{\cal F}_k(\phi)\s a^n, \label{eaction}\en
                   where the last sum on $k^{*}$, here made explicit, is a sum over nearest neighbors  in a positive sense from the site $k$ in all $n$ lattice directions, and
in both expressions the counterterm ${\cal F}_k(\phi)$ is given in (\ref{eF}). When one studies the
full action, as in a Monte Carlo analysis, then the small, artificial mass-like term can be omitted.

The generating function for Euclidean lattice spacetime averages is given, as usual, by
   \bn \< e^{\t  Z^{-1/2}\Sigma_k h_k\s\p_k\s a^n/\hbar}\>\equiv {\widetilde{M}}\int
e^{\t  Z^{-1/2}\Sigma_k h_k\s\p_k\s a^n/\hbar-I(\p,a,\hbar)/\hbar}\,\Pi_kd\phi_k \;, \label{elattice}
 \en
where $Z$ is the field strength renormalization constant.

In the next section
we study the perturbation
analysis of (\ref{elattice}) and determine that: (i) the proper {\it field strength
renormalization} is given by $Z=N'^{-2}(qa)^{1-s}$, (ii) the proper {\it mass renormalization} is given by
  $m_0^2=N'(qa)^{-1}$ $m^2 $, and (iii) the proper {\it coupling constant renormalization} is given by
$\l_0=N'^3(qa)^{s-2}\s \l$.
  Here, $q$ denotes a positive constant with dimensions (Length)$^{-1}$ ($q=b^{1/s}$ is a possible choice), and $m$ and $\l$ represent finite physical factors. It is noteworthy that $Z\s m_0^2=m^2/[N'(qa)^s]$
and $Z^2\s \l_0=\l/[N'(qa)^s]$. This completes the characterization of the model. {\bf {(}Remark:} Although
we have confined attention to models with quartic interactions, measure mashing also enables higher powers,
e.g., $\varphi^{44}_n$, $\varphi^{444}_n$, etc., to be handled just as well with the same counterterm \cite{kla4}. Even though
we have not discussed mixed odd and even potentials, they can also be treated.{\bf)}

A few philosophical remarks are in order. Much has changed by passing from a free model to a pseudofree model as the center of focus. Traditionally, when
forming local products from free-field operators, normal ordering is used. On the contrary,
after measure mashing, the pseudofree field operators satisfy multiplicative renormalization, and no normal
ordering is involved. Indeed, the very coefficients $m_0^2$ and $g_0$ act partially as multiplicative
renormalization factors for the associated products involved. To say that there are no divergences means,
for example, that the expression $m_0^2\Sigma'_k\p_k^2\s a^s$ is well defined, and this fact is
established by ensuring that $m_0^2\s\Sigma'_k\<\p_k^2\>\s a^s\propto N'a^s<\infty$. The same holds true
for $\l_0\Sigma'_k \p_k^4\s a^s$, which is shown to be well defined by noting that
$\l_0\Sigma'_k \<\p_k^4\>\s a^s\propto N'a^s<\infty$. These quantities remain bounded even in the
continuum limit.

  \section{The Continuum Limit, and Finiteness of a\\ Perturbation Analysis}
      Before focusing on the limit $a\ra0$ and $L\ra\infty$, we\index{relativistic model!continuum limit}
      note several important facts about ground-state averages of the direction
      field variables $\{\eta_k\}$. First, we assume that such averages
      have two important symmetries: (i) averages of an odd number
      of $\eta_k$ variables vanish, i.e.,
      \bn \<\eta_{k_1}\cdots\eta_{k_{2p+1}}\>=0\;, \en
      and (ii) such averages are invariant under any spacetime translation, i.e.,
    \bn
    \<\eta_{k_1}\cdots\eta_{k_{2p}}\>=\<\eta_{k_1+l}\cdots\eta_{k_{2p}+l}\>\;\en
    for any $l\in{\mathbb Z}^n$ due to a similar translational
    invariance of the lattice action. Second, we note that
    for any ground-state distribution, it is
    necessary that $\<\s\eta_k^2\s\>=1/N'$
for the simple reason that $\Sigma'_k\s\eta_k^2=1$. Hence,
       $|\<\eta_k\s\eta_l\>|\le1/N'$ as follows from the Schwarz
       inequality. Since $\<\s[\s\Sigma'_k\s\eta_k^2\s]^2\>=1$, it
       follows that $\<\s\eta_k^2\s\eta_l^2\s\>=O(1/N'^{2})$.
       Similar arguments show that for any ground-state distribution
         \bn  \<\eta_{k_1}\cdots\eta_{k_{2p}}\>=O(1/N'^{p})\;, \en
         which will be useful in the sequel.

         Our strategy for establishing that certain spacetime averages are finite relies on the
         discussion in Sec.~2.1 in
         which we showed that if the spatial average of a given quantity is finite it follows that
         the associated spacetime average is also finite. If the reader has forgotten how that argument
         goes, a review of Sec.~2.3 at this point may be useful.

         The continuum limit of the spatial lattice involves letting $L\ra\infty$ and $a\ra0$
         such that $La$ remains finite. It is of interest to compare this limit to one where
         $a$ is fixed and finite and $L\ra\infty$, as would be appropriate to a fixed lattice
         that grew to infinite size. This latter limit we call a thermodynamic limit instead of
         a continuum limit. These two limits are not the same, and we can exploit this difference to our
         advantage. In particular, we shall arrange matters so as to isolate all aspects of the
         lattice spacing $a$ (along with some terms involving $N'=L^s$) as a factor outside an integral
         in which only factors of $N'$ (and none involving $a$---in any essential way) reside inside the integral. Convergence of the factors outside the integral involves the continuum limit, while convergence of the integral
         reduces to that of a thermodynamic limit. For the integral, we shall assume that the unknown
         function $W$ may well effect the numerical value of the integral, but it will not effect whether
         or not the integral itself converges, that aspect being well covered by the large field behavior
         involving the $A_{k-l}$ terms as well as the small field behavior involving the $J_{k,l}$ terms.
\subsection{Field Strength Renormalization}
         For a suitable spatial test sequence $\{h_k\}$, we insist
         that expressions such as
         \bn \int Z^{-p}\,[\Sigma'_k h_k\s\p_k\,a^s]^{2p}\,\Psi_{pf}(\p)^2\,\Pi'_k\s
         d\p_k \label{w20}\en
         are finite in the continuum limit. Due to the intermediate
         field relevance of the factor $W$ in the pseudofree ground state,
         an approximate analysis of the integral will
         be adequate for our purposes.
          Thus, we are led to consider
         \bn &&\hskip-.3cm K\int Z^{-p}\,[\Sigma'_k
         h_k\s\p_k\,a^s]^{2p}\,\frac{e^{\t-\Sigma'_{k,l}\s\p_k\s A_{k-l}\s\p_l\,
         a^{2s}/\hbar-W}}{\Pi'_k[\s\Sigma'_lJ_{k,l}\p_l^2\s]^{(1-2ba^s)/2}}\,\Pi'_k\s
         d\p_k\no\\
         &&\hskip.2cm\simeq 2\s K_0\int Z^{-p}\s\k^{2p}\,[\Sigma'_k h_k\s\eta_k\,a^s]^{2p}\\&&\hskip1.4cm\times
         \frac{e^{\t-\k^2\s\Sigma'_{k,l}\s\eta_k\s A_{k-l}\s\eta_l\,a^{2s}/\hbar}\,\k^{R-1}}
         {\Pi'_k[\s\Sigma'_l J_{k,l}\s\eta^2_l\s]^{(1-2ba^s)/2}}\,d\k\,\delta(1-\Sigma'_k\eta_k^2)
         \,\Pi'_k\s d\eta_k\;, \label{f5} \no\en
         where we set $K_0$ as the normalization factor when $W$ is dropped.
         Our  goal is to use this integral to determine a value for
         the field strength renormalization constant $Z$.
         To estimate this integral we first replace two
         factors with $\eta$ variables by their appropriate
         averages. In particular, the quadratic expression in the exponent
         is estimated by
          \bn \k^2\s\Sigma'_{k,l}\s\eta_k\s
          A_{k-l}\s\eta_l\,a^{2s}\simeq\k^2\s\Sigma'_{k,l}\s N'^{\,-1}
          A_{k-l}\,a^{2s}\simeq\k^2\s N'\s a^{2s}\s a^{-(s+1)}\;,\en
          using the fact that the order of magnitude of the matrix $A_{k-l}\s\s a^{(s+1)}=O(1)$ \cite{nbook}.
          Next, the expression in the integrand is estimated by
          \bn [\Sigma'_k h_k\s\eta_k\,a^s]^{2p}\simeq\s
          N'^{\,-p}\,[\Sigma'_k h_k\,a^{s}]^{2p}\;. \en
         The integral over $\k$ is then approached by first rescaling the variable
         $\k^2\ra\k^2/(N'\s a^{s-1}/\hbar)$, which then leads to an overall integral
         estimate proportional to the coefficient
         \bn  Z^{-p}\,N'^{-p}\,[\Sigma'_k h_k\,a^{s}]^{2p}/[N'\s a^{s-1}]^p\;.\en
         At this point, all essential factors of $a$ are now outside the
         integral; indeed $R=2ba^sN'$ is assumed to be fixed and finite, and the factor $2ba^s$ that
          appears in the denominator can be taken to vanish without any change in the convergence
          properties. [We also note that had we kept the term $W=W(\p\s a^{(s-1)/2}/\hbar^{1/2})$ in our
     calculations, it too would no longer depend on the lattice spacing $a$ after the latest change
of variables.] For this final result to be meaningful in the continuum limit,
         we are led to choose $Z=N'^{\,-2}\s a^{-(s-1)}$. However, $Z$ must
         be dimensionless, so we introduce a fixed positive quantity
         $q$ with dimensions of an inverse length, which allows us to
         set
          \bn Z=N'^{\,-2}\s (q\s a)^{-(s-1)}\;; \en
          as noted above, we may choose $q=b^{1/s}$ if so desired.
 \subsection{Mass and Coupling Constant Renormalization}
 A power series expansion of the mass and coupling constant terms
 leads to the expressions
  $ \<\s [\s  m_0^2\,\Sigma_k \p_k^2 a^n\s]^p\s\> $ and
  $ \<\s [\s  \l_0\,\Sigma_k \p_k^4 a^n\s]^p\s\> $ for $p\ge1$, which we treat
  together as part of the larger family governed by
  $\<\s [\s  g_{0,r}\,\Sigma_k \p_k^{2r} a^n\s]^p\s\>$ for integral $r\ge1$.
  Thus we consider
  \bn &&\hskip-.2cm K\int [\s g_{0,r}\Sigma'_k
         \p_k^{2\s r}\,a^s]^{p}\,\frac{e^{\t-\Sigma'_{k,l}\s\p_k\s A_{k-l}\s\p_l\,
         a^{2s}/\hbar-W}}{\Pi'_k[\s\Sigma'_lJ_{k,l}\p_l^2\s]^{(1-2ba^s)/2}}\,\Pi'_k\s
         d\p_k\no\\
         &&\hskip.7cm\simeq 2\s K_0\int g_{0,r}^p\s\k^{2 r p}\,[\Sigma'_k\s\eta_k^{2 r}\,a^s]^{p}\label{ff5}\\
         &&\hskip1.7cm\times\frac{e^{\t-\k^2\s\Sigma'_{k,l}\s\eta_k\s A_{k-l}\s\eta_l\,a^{2s}/\hbar}\k^{R-1}}
         {\Pi'_k[\s\Sigma'_l J_{k,l}\s\eta^2_l\s]^{(1-2ba^s)/2}}\,d\k\,
         \delta(1-\Sigma'_k\eta_k^2)
         \,\Pi'_k\s d\eta_k\;.\no  \en
         The quadratic exponent is again estimated as
         \bn \k^2\s\Sigma'_{k,l}\s\eta_k\s
          A_{k-l}\s\eta_l\,a^{2s}\simeq \k^2\s N'\s a^{2s}\s a^{-(s+1)}\;, \en
while the integrand factor
         \bn [\Sigma'_k\eta_k^{2r}]^p\simeq N'^p\s N'^{-rp}\;. \en
        The same transformation of variables used above precedes the integral
        over $\k$, and the result is an integral, no longer depending on $a$ in any essential way, that is proportional to
          \bn g_{0,r}^p \s N'^{-(r-1)p}\s a^{sp}/[N'^{r}\s a^{(s-1)r}]^p\;.\en
          To have an acceptable continuum limit, it suffices that
              \bn g_{0,r}=N'^{(2r-1)}\,(q\s a)^{(s-1)r-s}\,g_r \;,\en
              where $g_r$ may be called the physical coupling factor.
              Moreover, it is noteworthy that
              $Z^r\,g_{0,r}=[N'\s (q\s a)^s]^{-1}\,g_r $,
          for all values of $r$, and which for a finite spatial volume $V'=N'\s a^s$ leads to a
    finite nonzero result for $Z^r\s g_{0,r}$. It should not be a surprise that there are no
          divergences for all such interactions because the source of all divergences has been
          neutralized!

    We may specialize the general result established above to the two cases of
    interest to us. Namely, when $r=1$ this last
    relation implies that $m_0^2=N'\s(q\s a)^{-1}\,m^2$, while when
    $r=2$, it follows that $\l_0=N'^{\,3}\s(q\s a)^{s-2}\s\s\l$. In
    these cases it also follows that $Z\s m_0^2=[\s N'\s (q\s a)^s\s]^{-1}\s m^2$ and
  $Z^2\s \l_0=[\s N'\s (q\s a)^s\s]^{-1}\s \l$,
    which for a finite spatial volume $V'=N'\s a^s$ leads to a
    finite nonzero result for $Z\s m_0^2$ and $Z^2\s \l_0$, respectively. Expressions
    for $Z$, $m_0^2$, and $\l_0$ provide
    sufficient information to fully define the model.

\section{Extension to Less Singular Scalar Models}
Let us take up the extension of measure mashing to other models such as $\varphi^4_n$, for  $n\le4$.
Although the classical pseudofree theory is identical to the classical free theory in these
cases, this fact does not prevent us from suggesting the consideration of mashing the measure for
such less singular models in an effort to eliminate divergences that arise in those cases. For $n=2$,
it is well known that normal ordering removes all divergences, but it is also well known that normal
ordering is a rather strange rule to define local products. In particular, if we rewrite the
product of two free-field operators as
\bn \varphi(x)\s\varphi(y)
=\<0|\s\varphi(x)\s\varphi(y)\s|0\>+:\varphi(x)\s\varphi(y): \;,\en
then, as $y\ra x$, the most singular term on the right-hand side is the first term, but since it is a multiple of unity, the {\it second
and less singular term} is chosen to define the local product, $ \varphi(x)^2_{Renormalized}=\;\,:\varphi(x)^2:$; moreover, this expression is {\it not} positive despite being the chosen local ``square'' of the field.
In sharp contrast, in the operator  product expansion, schematically given by
  \bn \varphi(x)\s\varphi(y)= c_1(x,y)\,\zeta_1(\half(x+y))
  +c_2(x,y)\,\zeta_2(\half(x+y))+\cdots\;, \en
the local product, as $y\ra x$, is defined as, say,
$\varphi(x)^2_{Renormalized}=\zeta_1(x)$, for which the
associated, dimensionless, $c$-number
coefficient $c_1(x,y)$ is the {\it most singular} as $y\ra x$; this is a very reasonable rule, and this choice is also positive as a square should be. To adopt measure mashing for $\phi^4_2$ would
introduce the operator product expansion and thereby a more natural local product definition. This same feature
also applies to $\p^4_3$,
and moreover it would also allow a self-consistent quantization of a model, in a three-dimensional spacetime, with the classical action
\bn  A=\tint(\half\s\{{\dot\phi}(x)^2-[{{\nabla}\phi}(x)]^2-m^2_0\s\phi(x)^2\s\} -\l_0\s\phi(x)^4\s
-g_0\s\p(x)^8)
\,d^3\!x\;, \en
with the aim of obtaining a faithful quantization of this combined system, namely, one for which
the classical limit of the quantized theory yields the original classical model; observe that this model includes both a super renormalizable {\it and} a nonrenormalizable interaction,. Measure mashing can also
be applied to $\p^4_4$ leading to a nontrivial proposal for this model, which is widely believed to be trivial when quantized conventionally; indeed, as already noted in the Introduction, preliminary Monte Carlo data  suggest a nontrivial behavior
for the $\p^4_4$ model when the special counterterm is included in the analysis \cite{stank}.
Finally, we observe that the extension of measure mashing to low-dimensional models would eliminate divergences that appear in such models when quantized conventionally, a feature that would certainly appear to be desirable.

\subsection{Extension to Higgs-like Fields}
The extension of our analysis to multi-component scalar fields, such as arise in the Higgs model as it appears
in the standard model in high energy physics, is quite straightforward. Assuming a natural rotational symmetry
among the separate fields, the basic requirement for the
proposed pseudofree model involves replacing the terms $\Sigma_{k,l}\p_k\s A_{k-l}\s \p_l$ by
$\Sigma_{k,l,\a}\p_{k,\a}\s A_{k-l}\s \p_{l,\a}$, and $\Sigma_k \p^4_k\s $ by $\Sigma_k [\Sigma_\a\p^2_{k,\a}]^2$,
where $\a\in\{1,2,...,A\}$ for an $A$-field scalar multiplet.
Likewise, the factor $\Pi'_k[\Sigma'_l J_{k,l}\p_l^2]^{-(1-R/N')/2}$ is replaced by
$\Pi'_k[\Sigma'_{l,\a} J_{k,l}\p_{l,\a}^2]^{-(1-R/N')/2}$. Finally, the hyperspherical coordinates for the
multi-field case are now taken to be $\p_{k,\a}=\k\s\eta_{k,\a}$, where $\k^2=\Sigma'_{k,\a}\s\p_{k,\a}^2$ and
$1=\Sigma'_{k,\a}\s\eta^2_{k,\a}$. It follow that $N'\equiv A\s L^s$, and it still appears reasonable to choose $R=2ba^sN'$. A few additional remarks about multi-component fields appear at the end of the Appendix.

\section{Commentary}
Is all this, including the nontraditional counterterm, a physically realistic proposal? Presumably, the answer depends on the application, so it
is too soon to expect a response to this question. Nevertheless,
it would seem there is some progress already just to have a possible solution to certain (non)renormalizable models
rather than the unsatisfactory results obtained by conventional techniques.

In the Appendix, we present a discussion of affine coherent states appropriate to the modified models
that has the promise of establishing that the classical limit of the continuum limit for an interacting
model is the original, nonlinear classical covariant field theory that was initially quantized.

\section{Acknowledgements}
The journey for the author to reach his present level of understanding of the quantization of
nonrenormalizable scalar quantum field theories has been long and has had its share of ups and downs.
Along the way, help and advice have been given by A.~Ahmed, E.~Deumens, J.~Stankowicz, and G.~Watson, and
they are thanked for their generous assistance. Suggestions for improvements to this article were made
by both S.V.~Shabanov and one of the referees to whom thanks are also given.

\section{Appendix: Affine Coherent States and the \\Quantum/Classical Connection}
To study the quantum/classical connection, it is very useful to introduce coherent states appropriate to the
system under consideration. As mentioned before, the basic operators associated with ground states that have undergone measure mashing, are,
in the continuum limit, affine field operators and not canonical operators.
Thus we introduce affine coherent states based on the affine operators and do so in a fashion so that
the operator representation that is chosen is compatible with the dynamics in the continuum limit. In order
to do so, it is traditional to use coherent states for which the fiducial vector is chosen as a vector
in the proper representation space, and one of the most useful ways to do this is to choose the fiducial vector to be the ground state of the Hamiltonian that contains the special counterterm that we have discussed previously.
How we use these states to relate the quantum theory to the classical theory is an application of a general strategy \cite{klaCRT23,wcp} that is most easily explained on simple, one-dimensional systems. Let us review that connection
next for two different types of simple systems.\v

{\it Canonical Variables:} For a one-dimensional system based on Heisenberg operators $Q$ and $P$ that obey
$[Q,P]=i\s\hbar\s\one$, we define the coherent states to be
  \bn  |p,q\>\equiv e^{\t-iqP/\hbar}\,e^{\t ip\s Q/\hbar}\,|\eta\>\;, \en
  where $|\eta\>$ is the fiducial unit vector. For simplicity, we assume that the fiducial vector
  is ``physically centered'', which means that $\<\eta|P|\eta\>=0$ and $\<\eta|Q|\eta\>=0$, a modest restriction
  on the fiducial vector. In this case it follows that $\<p,q|P|p,q\>=p$ and $\<p,q|Q|p,q\>=q$, and thus the physical significance of
  the $c$-numbers $p$ and $q$ is that of {\it mean values in the coherent states}; they decidedly are {\it not}
  sharp eigenvalues for either $P$ or $Q$. Additionally,  we assume that $|\eta\>$ is chosen so that
  $\<\eta|\s[\s P^2+Q^2\s]\s|\eta\>\ra0$ as $\hbar\ra0$, and that all necessary domain conditions hold.

  Next, recall that Schr\"odinger's equation may be derived from
  an abstract variational principle given---the subscript $Q$ stands for quantum--- by
    \bn I_Q=\tint\<\psi(t)|[\s i\hbar\d/\d t-\H(P,Q)\s]|\psi(t)\>\,dt\;, \en
    which under independent variations of $\<\psi(t)|$ and $|\psi(t)\>$ lead to Schr\"odinger's equation,
    \bn i\s\hbar\d\,|\psi(t)\>/\d t=\H(P,Q)\s|\psi(t)\>\;, \en
    along with its adjoint.  This is the quantum side of the variational principle.

   Now consider a {\it macro}scopic experimenter limited in her study of a {\it micro}scopic one-degree of freedom
   system so that she is vastly restricted in the actual variations of $|\psi(t)\>$ that she is able to make.
   Indeed, let us assume she can only move the system to a different location and/or change its velocity
   by a constant amount [N.~B.: We say ``velocity'' noting that $p\equiv\d\s L({\dot q},q)/\d{\dot q}\equiv p({\dot q},q)$ (with $L$ being the Lagrangian),
   and thus the velocity ${\dot q}={\dot q}(p,q)$]. The experimentalist is unable to probe the system at the
   microscopic level so that  she  can not
   make any changes to the state vector $|\psi(t)\>$ other than those regarding location and
   velocity.
   In stating these restrictions, we have limited the experimenter to the variational set of states for which
   $|\psi(t)\>\ra|p(t),q(t)\>$ for some fixed fiducial vector, namely, we have limited her just to the set of coherent states. In this
   case it follows that
   \bn I_Q\ra I_{Q\,restricted}\equiv\tint\<p(t),q(t)|[\s i\hbar\d/\d t-\H(P,Q)\s]|p(t),q(t)\>\,dt\;,\en
   which readily leads to
      \bn I_{Q\,restricted}\equiv\tint[p(t){\dot q}(t)-H(p(t),q(t))\s]\,dt\;, \en
      an expression that has all the appearance of being the action functional for a classical system, and the
      stationary variation of which, accounting for the proper boundary conditions, leads to the equations
      \bn {\dot q}(t)=\d\s  H(p,q)/\d p(t)\;,\hskip2em {\dot p}(t)=-\d\s H(p,q)/\d q(t)\;, \en
      two equations that have all the appearance of being Hamilton's dynamical equations of motion.

      Let us examine
      this alleged relationship with a classical theory more closely. For one thing, the
      proposed Hamiltonian $H(p,q)$ is given by
  \bn H(p,q)\hskip-1.2em&&\equiv\<p,q|\H(P,Q)|p,q\>\no\\&&=\<\eta|\H(P+p,Q+q)|\eta\>\no\\
             &&=\H(p,q)+\<\eta|[\s\H(P+p,Q+q)-\H(p,q)\s]|\eta\>\;. \label{you}\en
  In the last line of this expression, and apart from the first term $\H(p,q)$, the second term is $O(\hbar;p,q)$ so that in the limit $\hbar\ra0$,
  we find that $H(p,q)=\H(p,q)$. In short, the classical-looking system that has arisen from the restricted version
  of the quantum action functional is {\it the very classical system associated with the given quantum
  system}. There is only one additional point to clarify. Normally, we say for a classical system that the variables
  $p$ and $q$ are the {\it exact values} of the momentum and position, implying that these values are absolutely sharp
  values in the classical view. On the other hand, before $\hbar\ra0$, the meaning of $p$ and $q$ is that of
  {\it mean values} and not of sharp values. In the world in which we all live, $\hbar$ is {\it not} zero and
  therefore the classical and quantum systems must {\it coexist}. Thus it makes sense to assert that the
  restricted quantum action functional which has the form of a classical system is in fact the correct
  action functional for the
  classical system associated with the given quantum system and that in fact---still referring to the real
  world---it is consistent to assume that the true classical variables are mean values of some other variables.
  After all, who has ever measured the classical values of $p$ and $q$ to, say, $10^{137}$ decimal places
  to verify that they really are the exact momentum and position as hypothesized? In summary, we are
  led to propose that {\it the restricted variational form of the quantum action
  functional is the true classical action functional and its limited variation leads to the true
  classical equations of motion}.

  The quantum corrections arising from the second term in ({\ref{you})
  may vary depending on different choices of the fiducial vector $|\eta\>$; this property simply reflects the fact
  that the restricted action functional involves a projection from a larger space, and different projections
  can lead to differing elements.  Of course, these quantum corrections
  are generally extremely tiny and almost always can be neglected; when that is the case, we may say that
  the resultant equations of motion are strictly classical with no dependence
  on $\hbar$ whatsoever. However, as one may imagine, there are some exceptional
  systems where these corrections play a significant qualitative role. Even when that is the case, these
  terms may just be nuisance factors that can be safely ignored, or they may act to change the physics
  in significant ways.\v

  {\it Affine Variables:} For this example, we choose a different set of basic operators, namely $Q$ and $D$,
  where $[\s Q,D\s]=i\hbar Q$, which is called the {\it affine commutation relation}. This equation is
  surprisingly close to the canonical commutation relation when we observe that $[ Q,P]Q=i\hbar Q$, which
  on bringing the extra $Q$ on the left side inside the commutator, leads to $[Q,D]=i\hbar Q$, with $D\equiv
  \half(QP+PQ)$. Clearly,  $D$ has the dimensions of $\hbar$. From a representation point of view,
  there is, up to unitary equivalence, just one
  inequivalent representation of the canonical commutation relation for self-adjoint operators, while there are
  three inequivalent
  representations of the affine commutation relation for self-adjoint operators, distinguished by the fact
  that $Q>0$, $Q<0$, and $Q=0$ (strictly speaking, all these uniqueness results apply to unitary operators generated by the self-adjoint operators). For classical systems for which $q>0$ is the physical realm, it is natural
  to use affine kinematical variables with elements $D$ and $Q>0$ since both of them can be realized as self-adjoint operators (in contrast to canonical operators). For systems where $-\infty<q<\infty$, it is necessary to combine representations, and for systems
   for which the classical Hamiltonian satisfies $H(p,q)=p^2/2m+V(q)$, where $V(-q)=V(q)$,  a convenient
  way to do so, from the coherent state point of view, is to adopt $D$ and $Q^2$ (rather than $Q$) as the
  fundamental operators. These operators also provide an irreducible representation of the affine
  commutation relations for which $[Q^2,D]=2i\hbar\s Q^2$. Dynamical systems with the symmetry assumed above
  yield energy  eigenvectors that are either even or odd under reflection, and we will see how this impacts our discussion
  of such systems. Although, $Q=0$ (or $q=0$) is not included, it appears as a set of
  measure zero in our study, and including it generally causes no essential problems.

   For all $(p,q)\in{\mathbb R}^2$ (except when $q=0$),
  we define the affine coherent states by the expression
     \bn |p,q\>\equiv e^{\t ip\s Q^2/2\s q\s\hbar}\,e^{\t -i\ln(|q|/\ell)D/\hbar}\s|\eta\>\;;\en
   here $\ell>0$ is a fixed factor that sets the scale and cancels the dimensions of $q$, and the fiducial unit vector $|\eta\>$ is
   chosen to be symmetric in the sense $\<-x|\eta\>=\<x|\eta\>$,
   where as usual $Q\s|x\>=x\s|x\>$. A moments reflection shows that all the coherent states share this symmetry, i.e.,
   $\<-x|p,q\>=\<x|p,q\>$, and thus these states only span the even subspace of $L^2({\mathbb R})$ functions.
   {\bf[Remark:} A similar set of coherent states based on an {\it anti}-symmetric fiducial vector can also be
   introduced to study the odd subspace of $L^2({\mathbb R})$ functions; however, we do not discuss this second
   set.{\bf]} For brevity, we set $\<\eta|(\cdot)|\eta\>\equiv\<(\cdot)\>$, and we
   restrict $|\eta\>$ so that $\<D\>=0$, $\<(DQ^2+Q^2D)\>=0$, and $\<Q^2\>=\ell^2$; hereafter, without loss of
   generality, we shall assume units are chosen so that $\ell=1$. Observe
   that in this formulation the coherent states $|-p,-q\>=|p,q\>$. We note further that
     \bn  e^{\t i\ln(|q|)D/\hbar}\s Q\s e^{\t -i\ln(|q|)D/\hbar}\hskip-1.3em&&=|q|\s Q\;, \no\\
     e^{\t i\ln(|q|)D/\hbar}\s P\s e^{\t -i\ln(|q|)D/\hbar}\hskip-1.3em&&= P/|q|\;, \no\\
      e^{\t -ip\s Q^2/2q\hbar}\s D\s e^{\t ip\s Q^2/2q\hbar}\hskip-1.3em&&= D+pQ^2/q \no\\
     e^{\t -ip\s Q^2/2q\hbar}\s P\s e^{\t ip\s Q^2/2q\hbar}\hskip-1.3em&&= P+pQ/q \;, \en
      and thus it follows that
       \bn \<p,q|Q^2|p,q\>\hskip-1.3em&&=q^2\<\eta|Q^2|\eta\>=q^2\;,\no\\
           \<p,q|D|p,q\>\hskip-1.3em&&=\<\eta|(D+pqQ^2)|\eta\>=pq\;,\no\\
           \<p,q|P^2|p,q\>\hskip-1.3em&&=\<\eta|(P/|q|+p|q|Q/q)^2|\eta\>=p^2+\<P^2\>/q^2\;,\en
      which implies that the physical meaning of $q^2$ and $pq$ are
      the mean values of $Q^2$ and $D$, respectively. Let us now apply these states.

      Just like the canonical case, we assert that the quantum action functional for affine variables is given by
      \bn I_Q=\tint\<\psi(t)|[i\hbar\d/\d t-\H(P,Q)]|\psi(t)\>\,dt\;, \en
      and its stationary variation leads to Schr\"odinger's equation
        \bn i\hbar\d\s|\psi(t)\>/\d t=\H(P,Q)\s|\psi(t)\> \en
        and its adjoint. We focus on Hamiltonians such that
           \bn \H(P,Q)=\half m^{-1}P^2+V(Q)\equiv \half m^{-1}P^2+\Sigma_{j=0}^J\s c_j\s Q^{2j} \;. \en

As with the canonical case, we now restrict our variations and consider
   \bn I_{Q\s restricted}=\tint\<p(t),q(t)|[i\hbar\d/\d t-\H(P,Q)]|p(t),q(t)\>\,dt \;,\en
   now for the affine coherent states. To proceed further, we first note that
           \bn i\hbar\<p,q|\s(\d/\d t)|p,q\>=\<\eta|\s[-\half({\dot{p/q}})\s q^2Q^2+(\dot{\ln|q|})D]|\eta\>=
           \half(p{\dot q}-q{\dot p})\;.\en
           For the Hamiltonian, we observe that
           \bn H(p,q)\hskip-1.3em&&\equiv\<p,q|[\s\half m^{-1}P^2+V(Q)\s]|p,q\>\no\\
                  &&=\<\eta|[\s\half m^{-1}(P/|q|+p|q|Q/q)^2+\Sigma_{j=0}^J\s c_j\s q^{2j} Q^{2j}\s]|\eta\>\no\\
                  &&=\half { m}^{-1} p^2+\Sigma_{j=0}^J\s c_j\s \<Q^{2j}\>\s q^{2j}+\half m^{-1}\<P^2\>\s q^{-2}\no\\
                  &&\equiv\half { m}^{-1} p^2+\Sigma_{j=0}^J\s c'_j\s q^{2j}+\half m^{-1}\<P^2\>\s q^{-2}\;, \en
                  where we have set
                  $c'_j\equiv c_j\s \<Q^{2j}\>= c_j+O(\hbar)$. In summary, it follows that
                  \bn I_{Q restricted}=\tint[\half(p{\dot q}-q{\dot p})-\half {m}^{-1}p^2-
                  \Sigma_{j=0}^J\s c'_j\s q^{2j}-\half m^{-1}{\tilde c}_{-1}\s\hbar^2\s q^{-2}\s]\,dt\;, \label{e62}\en
                  where we have also set ${\tilde c}_{-1}\equiv \<P^2\>/\hbar^2$, which, in the present units, is dimensionless.

                  Stationary variation of this action with respect to $p$ and $q$, subject to appropriate
                  boundary conditions, leads to the associated Hamilton equations of motion. Apart from
                  typically small $\hbar$ changes of the various constants, there appears an additional,
                  unexpected force (proportional to $\hbar^2$) that prohibits solutions from crossing
                  $q=0$. This situation qualitatively changes the solutions from those based on the true
                  classical theory in which we let $\hbar\ra0$ before deriving the equations of motion.
                  However, if we are permitted to take the strict classical limit $\hbar\ra0$ before deriving
                  the equations of motion by the variational principle, then our classical theory
                  would lead to all the expected solutions.

\subsection{Coherent States for Scalar Fields}
  By generalizing the one-dimensional example above based on affine coherent states, we now study the coherent states
  for covariant scalar fields; in this effort we are partially guided by an analogous story for ultralocal
  fields that appears in \cite{book} as well as a preliminary study of these questions
  in \cite{hindawi}.  We start with the lattice-regularized, covariant pseudofree theory, and we deliberately
  choose the ground state for this model as the fiducial vector. Thus we are led to consider (for $\hbar=1$)
  the set of states
  \bn &&\hskip-3em\<\p|p,q\>=K\s \Pi'_k \s|q_k|^{-1/2}
  \frac{e^{\t i\Sigma'_k (p_k/2q_k)\s \p_k^2\s a^s
  -\half\Sigma'_{k,l}(\p_k/|q_k|)\s A_{k-l}\s(\p_l/|q_l|)\s a^{2s}}}
  {\Pi'_k[\s\Sigma'_l\s J_{k,l}\s(\p_l^2/q_l^2)\s]^{(1-2ba^s)/4}}\no\\
        &&\hskip13em\times e^{\t -\half\s W(\s(\p/|q|)\s a^{(s-1)/2}/\hbar^{1/2})}\;. \en
 The coherent state overlap function $\<p',q'|p,q\>$ is given by
  \bn \<p',q'|p,q\>=\tint \<p',q'|\p\>\<\p|p,q\>\,\Pi'd\p_k\;, \en
  which is represented by
      \bn &&\hskip-3em\<p',q'|p,q\>=K^2\s\Pi'_k(|q'_k|\s|q_k|)^{-1/2}\no\\
      &&\hskip2em\times\int \frac{e^{\t -i\Sigma'_k (p'_k/2q'_k)\s \p_k^2\s a^s
  -\half\Sigma'_{k,l}(\p_k/|q'_k|)\s A_{k-l}\s(\p_l/|q'_l|)\s a^{2s}}}
  {\Pi'_k[\s\Sigma'_l\s J_{k,l}\s(\p_l^2/{q'_l}^2)\s]^{(1-2ba^s)/4}}\no\\
        &&\hskip9em\times e^{\t -\half\s W(\s(\p/|q'|)\s a^{(s-1)/2}/\hbar^{1/2})}\no\\
        &&\hskip3em\times\frac{e^{\t i\Sigma'_k (p_k/2q_k)\s \p_k^2\s a^s
  -\half\Sigma'_{k,l}(\p_k/|q_k|)\s A_{k-l}\s(\p_l/|q_l|)\s a^{2s}}}
  {\Pi'_k[\s\Sigma'_l\s J_{k,l}\s(\p_l^2/q_l^2)\s]^{(1-2ba^s)/4}}\no\\
        &&\hskip9em\times e^{\t -\half\s W(\s(\p/|q|)\s a^{(s-1)/2}/\hbar^{1/2})} \,\Pi'd\p_k\;. \en
        As we approach the continuum limit in this expression, and restricting attention to
        continuous functions $p_k\ra p({\bf x})$ and $q_k\ra q({\bf x})$, it follows that
        \bn &&\hskip-3em\<p',q'|p,q\>=K^2\s\Pi'_k(|q'_k|\s|q_k|)^{-ba^s}\no\\
      &&\hskip2em\times\int \frac{e^{\t -i\Sigma'_k (p'_k/2q'_k)\s \p_k^2\s a^s
  -\half\Sigma'_{k,l}(\p_k/|q'_k|)\s A_{k-l}\s(\p_l/|q'_l|)\s a^{2s}}}
  {\Pi'_k\s[\Sigma'_l\s J_{k,l}\s\p_l^2\s]^{(1-2ba^s)/4}}\no\\
        &&\hskip9em\times e^{\t -\half\s W(\s(\p/|q'|)\s a^{(s-1)/2}/\hbar^{1/2})}\no\\
        &&\hskip3.5em\times\frac{
        {e^{\t i\Sigma'_k (p_k/2q_k)\s \p_k^2\s a^s
  -\half\Sigma'_{k,l}(\p_k/|q_k|)\s A_{k-l}\s(\p_l/|q_l|)\s a^{2s}}}}
  {\Pi'_k[\s\Sigma'_l\s J_{k,l}\s\p_l^2\s]^{(1-2ba^s)/4}}\no\\
        &&\hskip9em\times e^{\t -\half\s W(\s(\p/|q'|)\s a^{(s-1)/2}/\hbar^{1/2})}\,\Pi'_kd\p_k\;,\label{eee}\en
        where we taken advantage of the fact that for continuous functions we can bring the
        $q'$ and $q$ factors out of the denominators in the former expression.
        Although we can not write an analytic expression for the entire continuum limit of this
        expression, we note that the new prefactor, $\Pi'_k(|q'_k|\s|q_k|)^{-ba^s}$, has a
        continuum limit given by
        \bn  \Pi'_k(|q'_k|\s|q_k|)^{-ba^s}\ra e^{\t-b\tint[\s\ln|q'({\bf x})|+\ln|q({\bf x})|]\,d{\bf x}}\;.
        \label{e86}\en

     {\it This meaningful partial result for the continuum limit holds only for the affine coherent states;
     it would decidedly not have led to meaningful results for canonical coherent states \cite{hindawi}. In other words, measure mashing has had the effect of changing a canonical system into an affine system!} This
     result also favors the choice $R=2ba^sN'$ as it is compatible with an infinite spatial volume.

        The rest of the coherent state overlap integral in (\ref{eee}) is too involved to be simplified, but if we ask only for
        $\<p',q|p,q\>$ then some progress can be made. In this case we have
  \bn \<p',q|p,q\>\hskip-1.3em&&=K^2\Pi'_k\s|q_k|^{-1}\no\\
  &&\hskip1em\times\int \frac{e^{\t -i\Sigma'_k (p'_k/2q_k)\s \p_k^2\s a^s
  -\half\Sigma'_{k,l}(\p_k/|q_k|)\s A_{k-l}\s(\p_l/|q_l|)\s a^{2s}}}
  {\Pi'_k[\s\Sigma'_l\s J_{k,l}\s(\p_l^2/{q_l}^2)\s]^{(1-2ba^s)/4}}\no\\
        &&\hskip5em\times e^{\t -W(\s(\p/|q|)\s a^{(s-1)/2}/\hbar)/2}\no\\
        &&\hskip3em\times\frac{e^{\t i\Sigma'_k (p_k/2q_k)\s \p_k^2\s a^s
  -\half\Sigma'_{k,l}(\p_k/|q_k|)\s A_{k-l}\s(\p_l/|q_l|)\s a^{2s}}}
  {\Pi'_k[\s\Sigma'_l\s J_{k,l}\s(\p_l^2/q_l^2)\s]^{(1-2ba^s)/4}}\no\\
        &&\hskip5em\times e^{\t -W(\s(\p/|q|)\s a^{(s-1)/2}/\hbar)/2} \,\Pi'd\p_k\no\\
        &&=K^2 \int \frac{e^{\t -i\Sigma'_k ((p'_k-p_k)q_k/2)\s \p_k^2\s a^s
  -\Sigma'_{k,l}\p_k\s A_{k-l}\s\p_l\s a^{2s}}}
  {\Pi'_k[\s\Sigma'_l\s J_{k,l}\s\p_l^2\s]^{(1-2ba^s)/2}}\no\\
       && \hskip5em\times e^{\t -W(\s\p\s\s a^{(s-1)/2}/\hbar)}\,\Pi'd\p_k \;.\label{e77} \en
       In this form, we see that $\<p',q|p,q\>=\<p'q,1|p\s\s q,1\>$,
       from which we learn
       that the expression $\<p'|p\>\equiv \<p',1|p,1\>$ contains the same information
       as  contained in $\<p',q|p,q\>$; we also learn that $\<p',q|p,q\>$ achieves a
       meaningful continuum limit provided that $\<p'|p\>$ already achieves one, and it is
       clear from the form of (\ref{e77}) that such a continuum limit holds.
       \subsection{Quantum/Classical Connection}
       In the same spirit as the single affine degree of freedom, we seek to connect the classical
       action functional with a restricted form of the quantum action functional. Thus we proceed
       directly to the expression
          \bn I_{Q\s restricted}=\tint[\<p(t),q(t)|\s[i\hbar(\d/\d t)-\H(P,Q)\s]|p(t),q(t)\>\,dt\;, \label{www}\en
          based on the Hamiltonian operator $\H$ for the lattice covariant pseudofree model and
          the associated coherent states. Following the calculation of the affine model, we are led to
          the expression (with $\hbar=1$, ${\hat\p}_k=\p_k$,  and assuming units are chosen so that $\<\eta|\p_k^2|\eta\>=\ell^2=1$) for (\ref{www}) given by
          \bn I_{Q\s restricted}\hskip-1.3em&&=\int\{\s\half\Sigma'_k (p_k{\dot q}_k-q_k{\dot p}_k)\s a^s-\<\eta|\s [\s\half \Sigma'_k(P_k/|q_k|+p_k|q_k|\p_k/q_k)^2\s a^s\no\\
          &&\hskip4em +\half\Sigma'_k(|q_{k^*}|\p_{k^*}-|q_k|\p_k)^2\s a^{s-2}+\half s(L^{-2s}a^{-2})\Sigma'_kq_k^2\p_k^2\s a^s
          \no\\&&\hskip4em +\half\Sigma'_k\F_k(|q|\s\p)\s a^s\s]|\eta\>-E_{pf}\}\,dt\no\\
          && =\int\{\half\Sigma'_k(p_k{\dot q}_k-q_k{\dot p}_k)\s a^s -\half\Sigma'_k( p_k^2 +\<P^2_k\>\s q_k^{-2})\s a^s\no\\ &&\hskip3em -\half\Sigma'_k\Sigma'_{k^*}\<\s(|q_{k^*}|\p_{k^*}-|q_k|\p_k)^2\>\s a^{s-2}-\half s(L^{-2s}a^{-2})\s \Sigma'_kq_k^2\s a^s
          \no\\&&\hskip3em -\half\Sigma'_k\<\F_k(\p)\>\s q_k^{-2}\s a^s-E_{pf}\}\,dt  \;, \en
          where in the last line we have made the sum over $k^*$ explicit.
          This equation has all the expected ingredients apart from the rather unusual term
          $\half\Sigma'_k\Sigma'_{k^*}\<\s(|q_{k^*}|\p_{k^*}-|q_k|\p_k)^2\>a^{s-2}$ that we investigate next. We observe that
            \bn &&\<\s(|q_{k^*}|\p_{k^*}-|q_k|\p_k)^2\>\no\\
            &&\hskip3em=q_{k^*}^2\<(\p_{k^*}-\p_k)^2\>+(|q_{k^*}|
            -|q_k|)^2\<\p_k^2\>\no\\
            &&\hskip8em+[(q_{k^*}^2-q^2_k)+(|q_{k^*}|-|q_k|)^2]\<(\p_{k^*}-\p_k)\p_k\>\no\\
            &&\hskip3em\equiv C_1\s q_{k^*}^2+(|q_{k^*}| -|q_k|)^2+C_2\s[(q_{k^*}^2-q^2_k)+(|q_{k^*}|-|q_k|)^2]\;,\no\\   \en
         where $C_j$, $j=1,2$, are constants due to translation invariance of the ground state.
         The first term contributes to the mass, the second term and the latter part of the third term
         contribute to the lattice derivative, $(1+C_2)(q_{k^*}-q_k)^2$, since for continuous functions
         the sign of $q_{k^*}$ and $q_k$ are identical except for the possible exception when they are both infinitesimal, in which
         case they make a negligible contribution to the sum.
         Moreover, in the continuum limit $C_2\ra0$, and compared to unity it may be omitted from that factor.
         The initial part of the last factor sums to
         zero thanks to the periodic boundary conditions. We observe that $\<P_k^2\>=C_3$ and
         $\<\F_k(\p)\>=C_4$ are also constants because of translation invariance. Combining the various separate terms all together leads to a restricted action that has all the expected ingredients. The same
         remarks about terms of the form $\Sigma'_k\hbar^2\s q_k^{-2}$ that held for the one-dimensional
         affine system also hold in the present case; for this example, it may be suggested to take the limit $\hbar\ra0$
         before deriving the classical equations of motion.

         The factor $E_{pf}$ that enters the expressions above may be eliminated in a very easy fashion.
         Since the fiducial vector $|\eta\>$ has in this case been chosen as the ground state for the
         pseudofree model, it follows that
           \bn \<p,1|\s\H_{pf}\s|p,1\>\hskip-1.3em&&=\<\s[\s\half\Sigma'_k(P_k+p_kQ_k)^2a^s+V(Q)-E_{pf}\s]\s\>\no\\
                &&=\half\Sigma'_k \s p^2_k\s a^s+\<[\s\half\Sigma'_k P_k^2\s a^s+V(Q)-E_{pf}\s]\s\>\no\\
                &&=\half\Sigma'_k\s p^2_k\s a^s\;, \en
                meaning that the final expression for the pseudofree
                restricted action (with $\hbar=1$) is given by
                \bn I_{Q\s restricted}\hskip-1.3em&&=\int\{\half\Sigma'_k(p_k{\dot q}_k-q_k{\dot p}_k)\s a^s -\half\Sigma'_k[ p_k^2 +C_3\s (q_k^{-2}-1)]\s a^s\no\\ &&\hskip1em -\half\Sigma'_k\Sigma'_{k^*}\s(q_{k^*}-q_k)^2\s a^{s-2}-\half s(L^{-2s}a^{-2})\s \Sigma'_k(q_k^2-1)\s a^s
          \no\\&&\hskip1em -\half\Sigma'_k C_4\s (q_k^{-2}-1)\s a^s\s\}\,dt\;.  \en

         If instead of the pseudofree model we dealt with an
         interacting model, then the ground state would have a different form as would the form of the Hamiltonian operator; the restricted action would reflect this difference by adding a nonlinear term
         to the functional form of the action.
         \subsection{Coherent States for Multi-Component Scalar \\(Higgs-like) Models}
         To deal with multi-component scalar fields we generalize the single-component scalar field introduced above. In particular, the field
         variables $Q_k\equiv\p_k\ra Q_{k,\a}\equiv\p_{k,\a}$ as well as  $P_k\ra P_{k,\a}$ defined so that
            \bn [Q_{k,\a},P_{l,\beta}]=i\hbar\s a^{-s}\s\delta_{k,l}\s\delta_{\a,\beta}\;. \en
         We also introduce  $D_{k,\a}=\half (P_{k,\a}\s Q_{k,\a}+Q_{k,\a}\s P_{k,\a})$ (no summation)
         which has the commutation properties
             \bn  [Q_{k,\a},D_{l,\beta}]=i\hbar a^{-s}\s\delta_{k,l}\s\delta_{\a,\beta}\s Q_{k,\a}\;.\en
          For the coherent states we adopt
           \bn |p,q\>\equiv e^{\t-i\Sigma'_{k,\a}\s p_{k,\a}\s Q^2_{k,\a}/(2q_{k,\a}\hbar)\s\s a^s}
           \,e^{\t i\Sigma'_{k,\a}\ln (|q_{k,\a}|/\ell)\,D_{k,\a}\s a^s/\hbar}\,|\eta\>\;. \en
           Hereafter, the study of the multi-component field case follows rather closely that of the
           single-component case.

\end{document}